\documentclass[11pt]{article}
\usepackage{amsfonts}
\usepackage[normalem]{ulem}
\usepackage[makeroom]{cancel}
\usepackage{soul}
\usepackage{amssymb}
\usepackage{amsmath}
\usepackage{epsfig}
\usepackage{color}
\usepackage{parskip}
\usepackage{graphicx, subfigure}
\newlength{\dinwidth}
\newlength{\dinmargin}
\newtheorem{theorem}{Theorem}

\newtheorem{proposition}{Proposition}

\newtheorem{remark}{Remark}
\newtheorem{lemma}{Lemma}
\newtheorem{example}{Example}

\def\i{{\bf i}}

\begin{document}

\title{Triangular Schlesinger systems and superelliptic curves}

\author{Vladimir Dragovi\'c$^1$, Renat Gontsov$^2$, Vasilisa Shramchenko$^3$}

\date{}

\maketitle

\footnotetext[1]{Department of Mathematical Sciences, University
of Texas at Dallas, 800 West Campbell Road, Richardson TX 75080,
USA. Mathematical Institute SANU, Kneza Mihaila 36, 11000
Belgrade, Serbia.  E-mail: {\tt
Vladimir.Dragovic@utdallas.edu} 
}

\footnotetext[2]{M.S. Pinsker Laboratory no.1, Institute for Information Transmission Problems of the Russian Academy of Sciences,
Bolshoy Karetny per. 19, build.1, Moscow 127051 Russia. E-mail: {\tt gontsovrr@gmail.com}
}

\footnotetext[3]{Department of mathematics, University of
Sherbrooke, 2500, boul. de l'Universit\'e,  J1K 2R1 Sherbrooke, Quebec, Canada. E-mail: {\tt Vasilisa.Shramchenko@Usherbrooke.ca}
}

\begin{abstract}

We study the Schlesinger system of partial differential equations in the case when the unknown matrices of arbitrary size $(p\times p)$
are triangular and the eigenvalues of each matrix form an arithmetic progression with a {\it rational} difference $q$,  the same for all matrices.
We show that such a system possesses a family of solutions expressed {\it via} periods of meromorphic differentials on the Riemann surfaces of
superelliptic curves. We determine the values of the difference $q$,  for which our solutions lead to explicit polynomial  or rational  solutions
of the Schlesinger system. As an application of the $(2\times2)$-case, we obtain explicit sequences of rational solutions and of one-parameter
families of rational solutions of Painlev\'e VI equations. Using similar methods, we provide algebraic solutions of particular Garnier systems.

\end{abstract}

\tableofcontents

\section{Introduction}

We consider the Schlesinger system \cite{Sch1912}
\begin{equation}\label{Schl}
dB^{(i)}=-\sum_{j=1, j\ne i}^N\frac{[B^{(i)}, B^{(j)}]}{a_i-a_j}\,d(a_i-a_j), \qquad i=1,\ldots,N,
\end{equation}
for $(p\times p)$-matrices $B^{(1)},\ldots, B^{(N)}$ depending on the variable $a=(a_1,\ldots,a_N)$ which
belongs to some disc $D$ of the space ${\mathbb C}^N\setminus\bigcup_{i\ne j}\{a_i=a_j\}$. Written in a PDEs form,
this becomes
\begin{equation}\label{hSchlesinger_intro}
\frac{\partial B^{(i)}}{\partial a_j} =
\frac{[B^{(i)}, B^{(j)}]}{a_i-a_j}\quad (i\ne j), \qquad\qquad  \frac{\partial B^{(i)}}{\partial a_i}=
-\sum_{j=1, j\neq i}^N\frac{[B^{(i)},B^{(j)}]}{a_i-a_j}.
\end{equation}
 These equations govern  an {\it isomonodromic} family of Fuchsian linear
differential systems
\begin{equation}\label{fuchs}
\frac{dy}{dz}=\biggl(\sum_{i=1}^N\frac{B^{(i)}(a)}{z-a_i}\biggr)y, \qquad y(z)\in{\mathbb C}^p,
\end{equation}
with varying singular points $a_1,\ldots,a_N$.
As follows from  the  isomonodromic nature of the Schlesinger system, the eigenvalues  $\beta_i^k$
of the matrices $B^{(i)}$ that solve this system are constant (see proof of Theorem 3 from  \cite{Bo}).  These eigenvalues
are called the {\it exponents} of the Schlesinger system and of the related  isomonodromic family (\ref{fuchs}) of Fuchsian systems,
at their varying singular points $z=a_i$.

As known, due to B.\,Malgrange \cite{Ma}, the Schlesinger system is {\it completely integrable} in $D$, that is, for any initial data $B^{(1)}_0,\ldots,B^{(N)}_0\in{\rm Mat}(p,{\mathbb C})$ and any $a^0\in D$,  it has the unique solution
$B^{(1)}(a),\ldots,B^{(N)}(a)$ such that $B^{(i)}(a^0)=B^{(i)}_0$, $i=1,\ldots,N$. Moreover, (the pull-backs of)
the matrix functions $B^{(i)}$ are continued meromorphically to the universal cover $Z$ of the space
${\mathbb C}^N\setminus\bigcup_{i\ne j}\{a_i=a_j\}$ and their polar locus $\Theta\subset Z$, called the {\it Malgrange divisor}, is described as a zero set of a function $\tau$, holomorphic on the whole space $Z$. Being locally descended to $D$, this global $\tau$-function, up to a holomorphic non-vanishing in $D$ factor, coincides with the local one satisfying
Miwa's formula \cite{JMU}
$$
d\ln\tau(a)=\frac12\sum_{i=1}^N\sum_{j=1, j\ne i}^N\frac{{\rm tr}
(B^{(i)}(a)B^{(j)}(a))}{a_i-a_j}\,d(a_i-a_j).
$$

In the present paper we are going to focus on upper triangular matrix solutions $B^{(i)}=(b_i^{kl})_{1\leqslant k,l\leqslant p}$, that is on those with $b_i^{kl}=0$ for $k>l$, with specific arithmetic restrictions on the exponents. Triangular solutions of the Schlesinger system are those and only those with triangular initial data, since any set of $N$ triangular matrices evaluating with respect to this system remains triangular, due to the form of the system.  Note that the exponents in this case coincide with the diagonal  entries: $\beta_i^k=b_i^{kk}$.

Motivation for the problem we are going to consider comes from the basic $p=2, N=3$ case and classical algebraic geometry.  It is well known that in such a traceless triangular case, with $a_1=0, a_2=1, a_3=x$, the off-diagonal matrix element $b_1^{12}=b_1^{12}(x)$ of the matrix $B^{(1)}$ satisfies a hypergeometric equation:
\begin{equation*}
x(1-x)\,b_{1xx}^{12}+[ {\bf c}-({\bf a}+{\bf b}+1)  x]\,b_{1x}^{12}- {\bf ab} \, b_1^{12}=0,
\end{equation*}
where ${\bf a}=-2\sum_{j=1}^3\beta_j^1\,, \;\;{\bf b}=-2\beta_3^1\,,$ and  ${\bf c}=1-2(\beta_1^1 + \beta_3^1)$.
In the special case $(\beta_1^1, \beta_2^1, \beta_3^1)=(1/4, -1/4, -1/4)$, one recognizes
the classical Picard-Fuchs equation:
\begin{equation*}
x(1-x)\,b_{1xx}^{12}+(1-2x)\,b_{1x}^{12}-\frac{1}{4}\,b_1^{12}=0,
\end{equation*}
whose solutions are given by linear combinations of the periods of the differential $du/v$ on the elliptic curve
$$
v^2=u(u-1)(u-x)
$$
 (see for example \cite{Man} and \cite{Cl}, formula (2.25), p. 61).
Let us note that in this case $\beta_i^1-\beta_i^2=\pm 1/2$.

The last observation motivates us to consider the following particular case:
{\it each tuple $\{\beta_i^1,\ldots,\beta_i^p\}$ forms an arithmetic progression with the same {\bf rational}
difference $q=n/m$, where  $n\ne0$ and $m>0$  are coprime.} Generalizing the relationship with the Picard--Fuchs equations, we prove that
the corresponding triangular system (\ref{hSchlesinger_intro}) possesses a family of solutions having algebro-geometric nature, namely they are
expressed {\it via} periods of meromorphic differentials on the  Riemann surfaces $X_a$ of a (varying) algebraic plane curve of {\it superelliptic} type
$$
\hat{\Gamma}_a=\{(z,w)\in{\mathbb C}^2\mid w^m=(z-a_1)\ldots(z-a_N)\}.
$$
These expressions for the matrix entries $b_i^{kl}(a)$ are presented in Theorem \ref{thm} from Section 2.1.

Superelliptic curves are of much interest nowadays as well as some other related classes of curves, like $Z_m$  curves or
$(m, N)$-curves, see \cite{BeSi, BEL, EG, HiSh, Leg, MP, OU, PTSY, Sa, XZ, Zh} and references therein. There are some differences and ambiguity across the literature in definitions of these classes. For us, (following J.\, Sander, Yu.\,Zarhin and others) superelliptic curves are those which can be represented by an equation of the form:
$$
w^m = P_N (z),
$$
where $P_N$ is any polynomial of degree $N$. Due to the nature of the matter considered in the present paper, the zeros of $P_N$ are additionally assumed
to be simple, thus the superelliptic curves considered here are smooth in the affine part.  It is possible to extend the study to a more general case of superelliptic curves with singularities in the affine part, which we are going to address in a separate publication.

Note that triangular and, more generally, reducible, Schlesinger systems of arbitrary size $p$ were already studied by B.\,Dubrovin and M.\,Mazzocco in \cite{DM2}, where the main question was the following: when are solutions of one Schlesinger system for $N$ $(p\times p)$-matrices expressed {\it via} solutions of some other ``simpler'' Schlesinger systems of smaller matrix size or involving less than $N$ matrices? (See also some investigations of triangular Schlesinger systems in this context in the case of small dimensions $p=2$, $p=3$ in \cite{Go}, \cite{GL}.) However, there was no restriction imposed on the exponents, and thus there was no discussion of the integration of such systems in an explicit, in particular algebro-geometric, form. Nevertheless, it was mentioned that triangular solutions are expressed {\it via} solutions of Lauricella differential systems.  For the latter, there are already known representations  by integrals of multivalued functions over several kinds of chains in $\mathbb C$ (see, for example, \cite{MS} where the question of the linear independence of such integrals is also solved). We propose an alternative analysis based on the algebro-geometric approach which also helps to obtain some elementary expressions such as polynomial or rational ones, as we clarify below.  On the other hand, in the previous papers which provide particular algebro-geometric solutions to the Schlesinger system (\cite{DIKZ}, \cite{KK}, \cite{DS1} for $p=2$, and \cite {EG}, \cite{Kor} for an arbitrary $p$ in the case of quasi-permutation monodromy matrices of the family (\ref{fuchs})) the specific character of the triangular case has not been taken into consideration. The first article on {\it triangular} algebro-geometric solutions of Schlesinger systems (in the case $p=2$) is the recent \cite{DS2}, where the hyperelliptic case $m=2$ is studied, and our present work is an improvement and extension of the latter.

Concluding our introduction, let us be more specific and describe in general some features of the proposed algebro-geometric approach. In the case of $n>0$ and  when $m$ and $N$ are coprime, the mentioned meromorphic differentials have only one pole, therefore are all of the second kind, {\it i.~e.} have no residues. Thus their integration over elements of the homology group $H_1(X_a,\mathbb Z)$ is well defined. The first main result of this paper is Theorem 1 in Section 2.1 which provides families of algebro-geometric solutions of the system (2). Theorem 2 from Section 2.3 answers a delicate question about the dimension of the families of the solutions obtained in Theorem 1.

As observed in Theorem 1  in the case when  $n$ is positive and  the greatest common divisor of $m$ and $N$ is bigger than 1, denoted $(m, N)>1$, or when $n$ is negative,  the involved meromorphic differentials have several
poles $P_1,\ldots,P_s$ and are of the third kind in general, {\it i.~e.} have non-zero residues, therefore one should use elements of $H_1(X_a\setminus\{P_1,\ldots,P_s\},\mathbb Z)$ to integrate them correctly. We observe another effect in this case: taking small loops encircling the poles of the differentials, one expresses the matrix entries $b_i^{kl}(a)$ {\it via} the residues of the differentials, which turn out to be polynomials or rational functions
in the variables $a_1,\ldots,a_N$.  These are the results  of Theorem \ref{thm_2} in Section \ref{sect_thm2} for $n$ positive  and of Theorem 4 from Section 3.2 for $n$ negative.

As a consequence of Theorem \ref{thm_2}, we calculate explicitly a rational solution of the Painlev\'e VI equation with the parameters
$$
\alpha=\frac{(n+1)^2}2, \quad \beta=-\frac{n^2}{18}, \quad \gamma=\frac{n^2}{18}, \quad \delta=\frac{9-n^2}{18},
$$
for  each {\bf positive} integer $n$ not divisible by $3$,  additionally observing quite a regular asymptotic behaviour of its zeros and poles with respect to $n$ tending to infinity,  see Section 4.1, Theorem \ref{thmrat1} and Proposition \ref{th:zeros}.  In the same fashion, Theorem \ref{thm_rat2} from Section 4.2 gives a {\it one-parameter family} of rational solutions  of the Painlev\'e VI equation with the parameters
$$
\alpha=\frac{(3n+1)^2}2, \quad \beta=-\frac{n^2}{2}, \quad \gamma=\frac{n^2}{2}, \quad \delta=\frac{1-n^2}{2},
$$
for each {\bf negative} integer $n$. The last Section 5 is devoted to the applications to Garnier systems. Some  algebraic solutions of particular
Garnier systems are computed explicitly in  Section 5.1, Theorem \ref{thm_5}, and Section 5.2, Theorem \ref{thm_garnier}.

\section{An upper triangular Schlesinger system}

Let us note that the generally non-linear system (\ref{Schl}) in the case of triangular $(p\times p)$-matrices
$B^{(i)}$ splits into a  set of $p(p-1)/2$ inhomogeneous linear systems, each system has $N$ unknowns
$b_1^{kl}(a),\ldots,b_N^{kl}(a)$ with $k,l$ fixed. Indeed, first for each fixed $k=1,\ldots,p-1$ one considers a homogeneous
linear system
\begin{eqnarray}\label{hom_syst}
d\,b_i^{k,k+1}(a) = -\sum_{j=1,j\ne i}^N \bigl(\beta_i^{k,k+1}b_j^{k,k+1}(a)-\beta_j^{k,k+1}b_i^{k,k+1}(a)\bigr)
\frac{d(a_i-a_j)}{a_i-a_j}, \\ \mbox{with\; }\beta_i^{k,k+1}=\beta_i^k-\beta_i^{k+1}, \quad\mbox{where}\quad  \beta_i^k=b_i^{kk}, \nonumber
\end{eqnarray}
with respect to the unknowns $b_1^{k,k+1}(a),\ldots,b_N^{k,k+1}(a)$. Written in a vector form for the vector
$$
b^{k,k+1}(a)=\bigl(b_1^{k,k+1}(a),\ldots,b_N^{k,k+1}(a)\bigr)^{\top}\in{\mathbb C}^N,
$$
this becomes a Jordan--Pochhammer system
$$
d\,b^{k,k+1}=\Omega\, b^{k,k+1},
$$
with the meromorphic (holomorphic in the disc $D$) coefficient matrix 1-form
$$
\Omega=\sum_{1\leqslant j<l\leqslant N} J_{jl}\,\frac{d(a_j-a_l)}{a_j-a_l},
$$
where $J_{jl}$ are constant $(N\times N)$-matrices. Each matrix $J_{jl}$ has only four non-zero entries: in the $j$-th row
the entry with the number $j$ is equal to $\beta_l^{k,k+1}$ while the entry with the number $l$ is equal to $-\beta_j^{k,k+1}$, and
in the $l$-th row the entry with the number $j$ is equal to $-\beta_l^{k,k+1}$ while the entry with the number $l$ is equal to
$\beta_j^{k,k+1}$ (see details in \cite{Le}). The Jordan--Pochhammer system is completely integrable (which, in particular, follows from the complete integrability of the Schlesinger system)  and thus the solution
space of this system is $N$-dimensional.

After solving  systems (\ref{hom_syst}) one can subsequently pass to considering the following inhomogeneous linear systems with respect to the
unknowns $b_1^{kl}(a),\ldots,b_N^{kl}(a)$, for each fixed pair $(k,l)$ with $l-k=2,3,\ldots,p-1$:
\begin{equation}\label{inhom_syst}
d\,b_i^{kl}(a) = -\sum_{j=1,j\ne i}^N \bigl(\beta_i^{kl}b_j^{kl}(a)-\beta_j^{kl}b_i^{kl}(a)\bigr)\frac{d(a_i-a_j)}{a_i-a_j}\,
+F_i^{kl}, \qquad \mbox{with\; }\beta_i^{kl}=\beta_i^k-\beta_i^l,
\end{equation}
where the inhomogeneity $F_i^{kl}=F_i^{kl}(b_j^{k,<l},b_j^{>k,l})$ is given by
\begin{equation}\label{inhom}
F_i^{kl}=-\sum_{j=1,j\ne i}^N\Bigl(\sum_{k<s<l}b_i^{ks}b_j^{sl}-\sum_{k<t<l}b_j^{kt}b_i^{tl}\Bigr)\frac{d(a_i-a_j)}{a_i-a_j}.
\end{equation}

A general property of triangular solutions of the Schlesinger system is their {\it holomorphic} continuability to the whole universal cover $Z$ of the space ${\mathbb C}^N\setminus\bigcup_{i\ne j}\{a_i=a_j\}$ or, equivalently, the absence of the Malgrange divisor for such solutions. This phenomenon may be explained either by the fact that solutions of linear differential systems, which the triangular Schlesinger system is reduced to, do not have any other singularities apart from the fixed ones,
$\bigcup_{i\ne j}\{a_i=a_j\}\subset{\mathbb C}^N$, or by Miwa's formula, which for a triangular solution looks like
$$
d\ln\tau(a)=\frac12\sum_{i=1}^N\sum_{j=1, j\ne i}^N
\frac{\alpha_{ij}}{a_i-a_j}\,d(a_i-a_j),
$$
where $\alpha_{ij}=\beta_i^1\beta_j^1+\ldots+\beta_i^p\beta_j^p$.
Thus $\tau(a)=\prod_{i<j}(a_i-a_j)^{\alpha_{ij}}$ is a non-zero
holomorphic function on the universal cover $Z$ of the space
${\mathbb C}^N\setminus\bigcup_{i\ne j}\{a_i=a_j\}$ and the Malgrange divisor is empty.

\subsection{A particular case of the exponents and solutions via periods}

Further we will concentrate on the case when all the differences $\beta_i^k-\beta_i^{k+1}$ are rational, $\beta_i^k-\beta_i^{k+1}=n/m$,  $m>0$,
with $n, m$  coprime, and are the same for all $i=1,\ldots,N$, $k=1,\ldots,p-1$.
This choice of $\beta_i^k-\beta_i^{k+1}$ leads to all systems
(\ref{hom_syst}) have the  same  form
\begin{equation}
\label{hom_rat}
d\,b_i^{k,k+1}(a) = -\frac nm\sum_{j=1,j\ne i}^N \bigl(b_j^{k,k+1}(a)-b_i^{k,k+1}(a)\bigr)\frac{d(a_i-a_j)}{a_i-a_j}, \qquad  i=1,\ldots,N.
\end{equation}
A similar simplification holds for each inhomogeneous system (\ref{inhom_syst}). Note that $\sum_{i=1}^Nd\,b_i^{k,k+1}(a)\equiv0$,
and thus system (\ref{hom_rat}) is equivalent to
\begin{eqnarray*}
&&\frac{\partial b_i^{k,k+1}}{\partial a_j} = -\frac nm\,\frac{b_i^{k,k+1} - b_j^{k,k+1}}{a_i-a_j}, \quad j\ne i,
\\
&&\sum_{i=1}^Nb_i^{k,k+1}={\rm const}.
\end{eqnarray*}

We show that in this particular case of the exponents the triangular Schlesinger system possesses a family of solutions expressed {\it via} periods of meromorphic differentials on the compact Riemann surface of the  non-singular  algebraic plane curve
$$
\{(z,w)\in{\mathbb C}^2\mid w^m=(z-a_1)\ldots(z-a_N)\}.
$$
In the case of $n>0$, this family depends on $(p-1-\nu)(N-1)$ parameters, where $\nu$ is the number of integers among $ 1, 2, \ldots, p-1$ that are
divisible by $m$; for negative $n$, the family depends on $(p-1)(N-1)$ parameters (see Remark \ref{rmk_2}).
Let us denote the corresponding projective curve by ${\Gamma}_a\subset{\mathbb CP}^2$. There are the following three cases:
\begin{itemize}
\item if $N> m$ we have
$$
{\Gamma}_a=\{(z:w:\lambda)\in{\mathbb CP}^2\mid w^m\lambda^{N-m}=(z-\lambda a_1)\ldots(z-\lambda a_N)\}
$$
with one point at infinity  $\infty=(0:1:0)\,;$
\item if $m>N$ we have
$$
{\Gamma}_a=\{(z:w:\lambda)\in{\mathbb CP}^2\mid w^m=\lambda^{m-N}(z-\lambda a_1)\ldots(z-\lambda a_N)\}
$$
with one point at infinity  $ \infty=(1:0:0)\,;$
\item if $m=N$ we have
$$
{\Gamma}_a=\{(z:w:\lambda)\in{\mathbb CP}^2\mid w^m=(z-\lambda a_1)\ldots(z-\lambda a_N)\}
$$
with $m$ points at infinity $\infty=\{(1:1:0)$, $(1:\varepsilon:0),\ldots,(1:\varepsilon^{m-1}:0)\}\,,$ where $\varepsilon=e^{2\pi{\bf i}/m}\,.$
\end{itemize}

The point at infinity is singular when $|m-N|>1$, and non-singular when $|m-N|=1$. In the special case $m=N$,
the points at infinity are non-singular.

By the well-known theorem on the resolution of singularities (see, for example \cite[\S7.1]{Kir}) there is a compact Riemann surface $X_a$
and a holomorphic mapping $\pi: X_a\rightarrow{\mathbb CP}^2$, whose image is ${\Gamma}_a$ and
$$
\pi: X_a\setminus\pi^{-1}(\{\infty\})\rightarrow{\Gamma}_a\setminus\{\infty\}
$$
is a biholomorphism. We introduce  differentials $\Omega_1^{(j)}(a),\ldots,\Omega_N^{(j)}(a)$ given on the affine part $\hat\Gamma_a$
of $\Gamma_a$ by:
\begin{equation}\label{formula_for_omega}
\Omega_i^{(j)}(a)=\frac{w^{jn} dz}{(z-a_i)}, \qquad i=1,\ldots,N, \quad j=1,\ldots,p-1.
\end{equation}
If $n>0$, these differentials are holomorphic on the affine part $\hat\Gamma_a$ of the curve.
 Their holomorphicity at the points $(a_i,0)\in  \hat\Gamma_a$  follows from the parametrization
$$
z=a_i+t^m, \qquad w=t\,O(1), \qquad t\rightarrow0,
$$
of  $\hat\Gamma_a$  near $(a_i,0)$. The pull-back $\pi^*\,\Omega_i^{(j)}$ of each $\Omega_i^{(j)}$ under the biholomorphic
mapping $\pi$ is a holomorphic differential on $X_a\setminus\pi^{-1}(\{\infty\})$, with poles at $\pi^{-1}(\{\infty\})\,.$

In the case $n<0\,,$  differentials $\Omega_i^{(j)}(a)$ have poles at the points  $(a_i,0)$ of  $\hat\Gamma_a\,.$ Their pull-backs $\pi^*\,\Omega_i^{(j)}$ have poles at $\pi^{-1}((a_i,0))$ for $i=1,\dots, N$ and vanish at $\pi^{-1}(\{\infty\})$  as we explain in the next section.

For simplicity of notation, we denote the pull-backs $\pi^*\,\Omega_i^{(j)}$ of the differentials again by $\Omega_i^{(j)}\,$, keeping in mind
the change of variables $\int_{\gamma}\pi^*\,\omega=\int_{\pi(\gamma)}\omega$ in a definite integral, and $\pi^{-1}((a_i,0))$ by $(a_i,0)$.

Now we formulate our main theorem.
\begin{theorem}
\label{thm}
Let the eigenvalues of each matrix $B^{(i)}$, $i=1,\ldots,N$, have the same rational difference: $\beta_i^k-\beta_i^{k+1}=n/m$ $(k=1,\ldots,p-1)$,
where  $n\in\mathbb Z$  and $m$ are coprime. If $n>0$ assume also that $m>1\,.$  Then the following triangular matrices $B^{(i)}=(b_i^{kl})$ satisfy system \eqref{hSchlesinger_intro}:
\begin{equation}\label{formula_for_b}
b_i^{kl}(a)=\oint_{\gamma_{l-k}}\Omega_i^{(l-k)}(a), \quad l>k,
\end{equation}
where $\gamma_1,\ldots,\gamma_{p-1}$ are arbitrary elements of
\begin{enumerate}
\item[\rm (a)] $H_1(X_a,\mathbb Z)$ if $m$, $N$ are coprime and $n>0$,
\item[\rm (b)] $H_1(X_a\setminus\pi^{-1}(\{\infty\}),\mathbb Z)$ if $m$, $N$ are not coprime and $n>0$.
\item[\rm (c)]  $H_1(X_a\setminus\{(a_1,0),\ldots,(a_N,0)\},{\mathbb Z})$ if $n<0\,.$
\end{enumerate}
$($These cycles do not depend on $a\in D$ if $D$ is sufficiently small.$)$
\end{theorem}

\begin{remark}\label{rmk_1}
{\rm In the case $n>0\,,$  we assume that $m>1$ because the case $m=1$ is trivial: the  differentials $\Omega_i^{(j)}$ are exact in that case and thus
the $B^{(i)}$ are constant  diagonal matrices.}
\end{remark}

Before proving Theorem \ref{thm} let us analyze how the local structure of the curve $\Gamma_a$ at its singular point at infinity depends on the
values of $m$ and $N$, and how the differentials $\Omega_i^{(j)}(a)$ behave near their poles, the points of the set $\pi^{-1}(\{\infty\})$.

\subsection{The local structure of $\Gamma_a$ at infinity}
\label{sect_local}

The implicit function theorem cannot give us a local parameter near the singular point of $\Gamma_a$, for this purpose one should consider
the Puiseux expansions at the point at infinity (using the Newton polygon of the curve, see details in \cite[\S\S 7.2, 7.3]{Kir}). Computing the Puiseux
expansions also allows us to determine the number of the points in the set $\pi^{-1}(\{\infty\})$. After doing this exercise we arrive to the following two cases, assuming $n$ to be positive.

\begin{enumerate}
\item[(a)] Let $N$ and $m$ be coprime. In this case the set $\pi^{-1}(\{\infty\})$ consists of one point $P$, hence the differentials $\Omega_i^{(j)}(a)$
have the only pole and are all of the second kind. That is why the integration is correctly defined along the elements of $H_1(X_a,\mathbb Z)$
in this case.

In a local parameter $t$ in a neighbourhood of the point $P\in X_a$, $t(P)=0$, the mapping $\pi: X_a\rightarrow\Gamma_a$ (the parametrization of
$\Gamma_a$) can be chosen to have the form
$$
z=1/t^m, \qquad w=\frac1{t^N}(1-a_1t^m)^{1/m}\ldots(1-a_Nt^m)^{1/m}.
$$
The genus $g(X_a)$ of the Riemann surface $X_a$ equals
$$
g(X_a)=\frac12(m-1)(N-1)
$$
in this case.

\item[(b)] Let $N$ and $m$ be not coprime, that is let there be an integer $s>1$ such that $N=sN_1\,,$ $m=sm_1\,,$ with coprime $N_1$ and $m_1$.
In this case the set $\pi^{-1}(\{\infty\})$ consists of $s$ points $P_1,\ldots,P_s$, and the differential $\Omega_i^{(j)}(a)$ has $s$ poles, one at each
of the points $P_1,\ldots,P_s$ at infinity, being of the third kind in general. Thus, for the integration of $\Omega_i^{(j)}(a)$  to be well-defined, one
uses the elements of $H_1(X_a\setminus\{P_1,\ldots,P_s\},\mathbb Z)$ as integration contours.

In a local parameter $t$ at each point $P_k\in\pi^{-1}(\{\infty\})$, $t(P_k)=0$, the mapping $\pi: X_a\rightarrow\Gamma_a$ (the parametrization of
$\Gamma_a$) can be chosen to have the form
$$
z=1/t^{m_1}, \qquad w=\frac{\varepsilon^{k-1}}{t^{N_1}}(1-a_1t^{m_1})^{1/m}\ldots(1-a_Nt^{m_1})^{1/m}, \quad \varepsilon=e^{2\pi{\bf i}/s},
$$
which  implies the coordinate representation of the differentials near the poles $P_k$, $k=1,\ldots,s$:
\begin{equation}\label{omegab}
\Omega_i^{(j)}=\frac{w^{jn} dz}{z-a_i}=\frac{\nu_k(1-a_1t^{m_1})^{jn/m}\ldots(1-a_Nt^{m_1})^{jn/m}}{t^{jnN_1+1}(1-a_it^{m_1})}\,dt,
\quad \nu_k=-m_1\,\varepsilon^{jn(k-1)}.
\end{equation}
The genus $g(X_a)$ of the Riemann surface $X_a$ equals
$$
g(X_a)=\frac12\bigl((m-1)(N-1)-s+1\bigr)
$$
in this case.
\end{enumerate}

Assuming $n$ to be negative, we see from (\ref{omegab}) that the differential $\Omega_i^{(j)}(a)$ vanishes
at the points $P_1,\ldots,P_s$ at infinity. In this case it has $N$ poles $(a_1,0),\ldots,(a_N,0)\in X_a$ and for the integration of $\Omega_i^{(j)}(a)$  to be well-defined, one uses the elements of $H_1(X_a\setminus\{(a_1,0),\ldots,(a_N,0)\},\mathbb Z)$ as integration contours.

\begin{remark}{\rm A non-singular case $N=m$ can be regarded as a particular case of (b), with $X_a=\Gamma_a$, $\{\infty\}=\{P_1,\ldots,P_m\}$,
and $s=N=m$, $N_1=m_1=1$.}
\end{remark}

\subsection{Proof of Theorem 1}

Note that for each fixed $i=1,\ldots,N$, the functions $b_i^{kl}$ with the same $l-k$, defined in Theorem \ref{thm}, coincide. As for every $s$ such that
$k<s<l$, there exists $t$ such that $k<t<l$ and $l-s=t-k$ (and hence $s-k=l-t$), the inhomogeneity (\ref{inhom}) of system (\ref{inhom_syst})
vanishes. Therefore suffices to prove that the functions $b_1^{kl},\ldots,b_N^{kl}$ satisfy (\ref{inhom_syst}) with $F_i^{kl}=0$:
\begin{equation*}
d\,b_i^{kl}(a) = -(l-k)\frac nm\sum_{j=1,j\ne i}^N \bigl(b_j^{kl}(a)-b_i^{kl}(a)\bigr)\frac{d(a_i-a_j)}{a_i-a_j},
\end{equation*}
or, written in an equivalent PDEs form,
\begin{eqnarray*}
&&\frac{\partial b_i^{kl}}{\partial a_j}=-(l-k)\frac nm\frac{b_i^{kl}-b_j^{kl}}{a_i-a_j}, \quad j\ne i, \\
&&\sum_{i=1}^Nb_i^{kl}={\rm const}.
\end{eqnarray*}

Differentiating the equality $w^m=P(z,a):=(z-a_1)\ldots(z-a_N)$ with respect to $a_j$, we obtain
\begin{equation*}
mw^{m-1}\frac{\partial w}{\partial a_j}=-\frac{P(z,a)}{z-a_j}
\end{equation*}
or, equivalently,
\begin{equation*}
\frac{\partial w}{\partial a_j}=-\frac1m\frac w{z-a_j}\,.
\end{equation*}

Thus  keeping in mind definitions (\ref{formula_for_b}) and (\ref{formula_for_omega}) of $b_i^{kl}$ and of $\Omega_i^{(l-k)}$,  for $j\ne i$ one has
\begin{eqnarray*}
\frac{\partial b_i^{kl}}{\partial a_j}&=&\oint_{\gamma_{l-k}}\frac{\partial\Omega_i^{(l-k)}(a)}{\partial a_j}=
-\frac nm(l-k)\oint_{\gamma_{l-k}}\frac{w^{(l-k)n} dz}{(z-a_i)(z-a_j)}=\\&=&
-\frac{n(l-k)}{m(a_i-a_j)}\oint_{\gamma_{l-k}}\Bigl(\frac1{z-a_i}-\frac1{z-a_j}\Bigr)w^{(l-k)n}dz=
-(l-k)\frac nm\frac{b_i^{kl}-b_j^{kl}}{a_i-a_j}.
\end{eqnarray*}
The proof of $\sum_{i=1}^N b_i^{kl}={\rm const}$ is also a straightforward computation: for every fixed $a$ there holds
\begin{equation*}
mw^{m-1}\,dw=\sum_{i=1}^N\frac{P(z,a)\,dz}{z-a_i}
\end{equation*}
and thus
\begin{equation*}
 m\,dw=\sum_{i=1}^N\frac{w\,dz}{z-a_i}\,.
\end{equation*}
Using this we obtain
\begin{equation*}
\sum_{i=1}^N b_i^{kl}=\oint_{\gamma_{l-k}}\sum_{i=1}^N\frac{w^{(l-k)n} dz}{(z-a_i)}=\frac m{(l-k)n}\oint_{\gamma_{l-k}} dw^{(l-k)n},
\end{equation*}
which is zero as an integral of an exact differential over a cycle. This proves Theorem \ref{thm}. {\hfill $\Box$}

\begin{remark}\label{rmk_1.5}
{\rm As explained in Section \ref{sect_local}, the number of independent contours  in the homology groups $H_1(X_a,\mathbb Z)$ or $H_1(X_a\setminus\pi^{-1}(\{\infty\}),\mathbb Z)$ is $L=(m-1)(N-1)=2g+s-1\,,$ where $s=(N, m)$ is the greatest common divisor of $m$ and $N$ and $g$
is the genus of the Riemann surface $X_a$:
\begin{itemize}
\item if $N$ and $m$ are coprime, then there are $L=2g$ basic cycles in $H_1(X_a,{\mathbb Z})\,;$
\item if $(N,m)=s>1$, then there are $s$ points $P_1,\dots, P_s$ in the set $\pi^{-1}(\{\infty\})$ and thus $L=2g+s-1$ basis cycles in
$H_1(X_a\setminus \{P_1,\dots, P_s\},{\mathbb Z})\,.$
\end{itemize}
In the homology group $H_1(X_a\setminus\{(a_1,0),\ldots,(a_N,0)\},{\mathbb Z})$, the number of generators is $L=(m-1)(N-1)+N-s=2g+N-1\,.$}
\end{remark}

Denoting $\mathcal A_1, \dots, \mathcal A_L$ generators of $H_1(X_a\setminus\pi^{-1}(\{\infty\}),\mathbb Z)$,   in the case $n>0\,,$ and generators of $H_1(X_a\setminus\{(a_1,0),\ldots,(a_N,0)\},{\mathbb Z})$ in the case $n<0\,,$  we see that Theorem \ref{thm} gives us the following family of solutions for $b_i^{kl}$ for each pair of indices $l>k\,:$ taking $\gamma_{l-k}=\sum_{j=1}^L c_j^{(l-k)} \mathcal A_j$ with $c_j^{(l-k)}\in\mathbb C$ we have
\begin{equation*}
b_i^{kl}(a) = \sum_{j=1}^L c_j^{(l-k)}\oint_{\mathcal A_j}\Omega_i^{(l-k)}(a)\;,\qquad c_j^{(l-k)}\in\mathbb C\,.
\end{equation*}
The number of independent parameters describing this family will be discussed in Section \ref{sect_independence}.
\medskip

\subsection {Linear independence of solutions}
\label{sect_independence}

Note that for each fixed pair $(k,l)$, $1\leqslant k<l\leqslant p$, the vector
$$
\bigl(b_1^{kl}(a),\ldots,b_N^{kl}(a)\bigr)^{\top}=\Bigl(\oint_{\gamma}\Omega_1^{(l-k)}(a),\ldots,\oint_{\gamma}\Omega_N^{(l-k)}(a)\Bigr)^{\top}
$$
is a solution of the Jordan--Pochhammer linear differential system of size $N$, where the cycle $\gamma$ belongs to $H_1(X_a,{\mathbb Z})$ or
$H_1(X_a\setminus\{P_1,\ldots,P_s\},{\mathbb Z})$ in the case of positive $n$ and to $H_1(X_a\setminus\{(a_1,0),\ldots,(a_N,0)\},{\mathbb Z})$ in the case of negative $n$. As $\sum_{i=1}^N b_i^{kl}=0$, the complete integrability of the latter system implies that
this vector belongs to an $(N-1)$-dimensional subspace of the $N$-dimensional solution space of the system. Thus it is natural to ask  whether
among the columns of the matrix
\begin{equation*}
{\bf B}(a)=\left(\begin{array}{ccc}\oint\limits_{\mathcal A_1}\Omega_1^{(l-k)}(a) & \dots & \oint\limits_{\mathcal A_{L}}\Omega_1^{(l-k)}(a)
 \\\vdots &  & \vdots \\ \\ \oint\limits_{\mathcal A_1}\Omega_N^{(l-k)}(a) & \dots & \oint\limits_{\mathcal A_{L}}\Omega_N^{(l-k)}(a)\end{array}\right),
\end{equation*}
there are $N-1$ linearly independent over $\mathbb C\,.$ In case the answer
is positive, we have an $(N-1)$-parameter family of algebro-geometric solutions of system (\ref{inhom_syst}).
 Here  if $n$ is positive, then  $L=(m-1)(N-1)=2g+s-1$ with $s=(N,m)\,,$ see Remark \ref{rmk_1.5}, and
the contours of integration
$\mathcal A_1,\ldots,\mathcal A_{L}$
form a set of generators of $H_1(X_a\setminus \{P_1,\dots, P_s\},{\mathbb Z})\,.$  In the case of negative $n$, we have $L=(m-1)(N-1)+N-s=2g+N-1$ and the contours $\mathcal A_1,\ldots,\mathcal A_{L}$ form a set of generators of the group  $H_1(X_a\setminus\{(a_1,0),\ldots,(a_N,0)\},{\mathbb Z})\,.$
\medskip

\begin{theorem}
\label{thm_conj}
 Let $n$ and $m$ be comprime.
If $n$ is positive and $l-k$ is not divisible by $m$ or if $n$ is negative,  then among the columns of the matrix ${\bf B}$ there are $N-1$ linearly independent over $\mathbb C$.
\end{theorem}

\begin{remark}\label{rmk_2}
{\rm Let  $n>0$ and $\nu$ be the number of integers among $1, 2, \ldots, p-1$ that are divisible by $m$. Then Theorem \ref{thm_conj} implies that
the algebro-geometric expressions of Theorem \ref{thm} generate a $(p-1-\nu)(N-1)$-parameter family of solutions of the triangular Schlesinger system (\ref{hSchlesinger_intro}) with fixed exponents as in Theorem \ref{thm}, whose solutions moduli space is of  dimension $p(p-1)(N-1)/2$. In the case $n<0\,,$ Theorem \ref{thm} yields a $(p-1)(N-1)$-parameter family of solutions to such  triangular Schlesinger system (\ref{hSchlesinger_intro}). In particular,
in the $(2\times2)$-case, $p=2$, {\it all} solutions of such a system are algebro-geometric.}
\end{remark}

{\it Proof of Theorem \ref{thm_conj}.}  First, let us denote $j=l-k$ and reformulate the statement of the theorem in the following  way:  Let $n$ and $m$ be comprime. If $n$ is positive and an integer $j$ is not divisible by $m$  or if $n$ is negative  then there exists $a^0\in D$ such that among the $N$ differentials
\begin{equation}
\label{Ndifferentials}
\Omega_i^{(j)}(a^0)=\frac{w^{jn} dz}{z-a_i^0}, \qquad i=1,\ldots,N,
\end{equation}
any $N-1$ differentials are linearly independent in the cohomology space $H^1(X_{a^0})$.
\medskip

Indeed, any $N-1$ columns of the matrix ${\bf B}$ are linearly dependent over $\mathbb C$ if and only if ${\rm rk}\,{\bf B}(a^0)<N-1$
for some $a^0\in D$ (this is due to that the columns of $\bf B$ are solutions of a completely integrable linear differential system). The latter
holds if and only if any $N-1$ rows of ${\bf B}(a^0)$ are linearly dependent, that is, a nontrivial linear combination of any $N-1$ differentials
among $\Omega_1^{(j)}(a^0),\ldots,\Omega_N^{(j)}(a^0)$ has all its periods equal to zero, which is equivalent to being an exact differential.

Let us now prove that among the differentials \eqref{Ndifferentials} any $N-1$ are  linearly independent. Suppose, on the contrary,  that there exist numbers $\alpha_1, \dots, \alpha_{N-1}\in\mathbb C$ such that the following linear combination
\begin{equation}
\label{phi}
\varphi = \sum_{i=1}^{N-1}\alpha_i\Omega_i^{(j)}(a)=\sum_{i=1}^{N-1}\alpha_i\frac{w^{jn} dz}{z-a_i}
\end{equation}
is an exact differential on the Riemann surface $X_a$ of the algebraic curve $w^m=P(z)$ (where $P(z)=(z-a_1)\ldots(z-a_N)$ with $a_1,\ldots,a_N$ fixed).

Denote $J:\hat\Gamma_a \to \hat\Gamma_a$ the symmetry of the underlying algebraic curve: $J(z,w) = (z,\varepsilon w)$ with $\varepsilon$ being an $m$th primitive  root of unity: $\varepsilon=e^{2\pi{\bf i}/m}$ and
consider separately the cases of positive and negative values of $n\,.$

\begin{itemize}
\item Let $n>0\,.$  In this case we assume that $j$ is not divisible by $m$.  The following integral of the exact differential $\varphi$
\begin{equation}
\label{y}
y(z,w) = \int_{(a_N,0)}^{(z,w)} \varphi
\end{equation}
is a well-defined  meromorphic  function on $X_a\,.$

We have
\begin{equation*}
J^*\varphi = \varepsilon^{jn}\varphi \qquad\mbox{and}\qquad J^*y = \int_{(a_N,0)}^{J(z,w)}\varphi =\int_{J(a_N,0)}^{J(z,w)}\varphi =   \int_{(a_N,0)}^{(z,w)}J^*\varphi = \varepsilon^{jn}y\,.
\end{equation*}

Let $k$ be the smallest integer such that $km-jn>0\,.$ If $m>jn$, then $k=1$, otherwise $k>1\,.$

The following  meromorphic  function on the surface $X_a$
\begin{equation}
\label{f1}
f= w^{km-jn}y
\end{equation}
is invariant under the symmetry $J$. Therefore it descends to a  meromorphic  function of $z$ defined on the base of the ramified covering $z:X_a\to\mathbb CP^1$.  Given that this function has the only pole at the point at infinity, we conclude that $f(z)$ is a polynomial.

Recall from Section \ref{sect_local} that $s=N/N_1=m/m_1$  and each differential $\Omega_i^{(j)}(a)$ has $s$ poles at points at infinity $P_1,\ldots,P_s\in X_a$ of order $jnN_1+1$ and that the local parameter at each of these points is $t=z^{-1/m_1}\,.$ Thus the differential $\varphi$ has poles at the points $P_1, \dots, P_s$  of order at most $jnN_1+1$ and the poles of function $y$ at those points are of order at most $jnN_1\,.$
Given that the function  $w$  has a pole of order  $N_1$  at each of the points $P_1, \dots, P_s$, we obtain that the poles of the function $f=w^{km-jn}y$ at the points $P_1, \dots, P_s$ are of the order at most $kmN_1\,.$ Therefore $f$ is a polynomial in $z$ of degree at most $kmN_1/m_1 = kN\,.$

On the other hand, the polynomial $f$ has $N$ zeros at $z=a_i$, $\;i=1,\dots, N\,$. Let us show that each zero is of multiplicity at least $k$.

Consider the function $J^*y$ evaluated at a branch point $(a_i,0)$ of the curve. On one hand, we know that $J^*y=\varepsilon^{jn} y$ and therefore
\begin{equation*}
J^*y(a_i,0) = \varepsilon^{jn}y(a_i,0)\,.
\end{equation*}
On the other hand, we have
\begin{equation*}
J^*y(a_i,0) = \int_{(a_N,0)}^{J(a_i,0)}\varphi = \int_{(a_N,0)}^{(a_i,0)}\varphi = y(a_i,0)\,.
\end{equation*}
The two above relations imply that $\varepsilon^{jn}y(a_i,0)=y(a_i,0)$ and, since $jn$ is not a multiple of $m$ (because $j$ is not, and $n$, $m$ are coprime), we conclude that $y(a_i,0)=0$ (note that $y$ does not have a pole at $(a_i,0)$ since $\varphi$ vanishes there).  The differential $dy = \varphi$ vanishes at $(a_i,0)$ to the order $jn-1$ (recall that the local parameter near the ramification point is $t_i=(z-a_i)^{1/m}$)\,. Thus we have that the function $y$ vanishes at every  finite  ramification point to the order $jn\,.$

Coming back to  $f(z)$ defined by \eqref{f1} and considered as function on the $z$-sphere, we find that  it behaves as $O((z-a_i)^k)$ at the branch point $z=a_i$ and thus it has $N$ zeros of order $k$ at $a_1,\dots, a_N\,.$ We can now conclude that $f(z)$ is proportional to $P^k(z)\;:$
\begin{equation*}
w^{km-jn}y = c\; P^k(z)
\end{equation*}
with some constant $c$ which may depend on the $\{a_i\}\,.$ From here we obtain $y=c\; P^k(z)w^{jn-km}= c\;w^{jn}$ and thus
\begin{equation}
\label{concl1}
\varphi=dy = c\; d w^{jn}\,.
\end{equation}

\item Let $n<0$ and assume that $j$ is not a multiple of $m_1\,.$ In this case the function $w^{jn}$ has a zero at each of the points $P_1,\dots, P_s$ and therefore
the differential $\varphi$ does. Define
\begin{equation}
\label{y2}
y(P) = \sum_{i=1}^s\int_{P_i}^{P} \varphi\;,
\end{equation}
which is a well-defined meromorphic function on $X_a\,,$ given that the differential $\varphi$ is exact.

The symmetry $J$ permutes the set of the points at infinity $\{P_1, \dots, P_s\}$ having the period $s$ on this set: $J^s(P_i)=P_i$, $i=1,\ldots,s$.
We have the following behaviour under the symmetry $J\;:$
\begin{equation*}
J^*\varphi = \varepsilon^{jn}\varphi \qquad\mbox{and}\qquad J^*y = \sum_{i=1}^s\int_{P_i}^{J(P)}\varphi =\sum_{i=1}^s\int_{J(P_i)}^{J(P)}\varphi =   \sum_{i=1}^s\int_{P_i}^{P}J^*\varphi = \varepsilon^{jn}y\,.
\end{equation*}
Let $k$ be the smallest integer such that $-jn-km<0$ and define the following  meromorphic function on the surface $X_a\,:$
\begin{equation}
\label{f2}
f= w^{-jn-km}y\,.
\end{equation}
Similarly to the previous case, this function is invariant under the symmetry $J$ and therefore descends to a meromorphic function of $z$ defined
on the base of the ramified covering $z:X_a\to\mathbb CP^1$, having now poles at $z=a_i$ with $ i=1,\dots, N$ and a zero at $z=\infty$.

The function $w^{-jn-km}$ has a pole of order $jn+km$ at $(a_i,0)$ (with respect to the local parameter $t_i=(z-a_i)^{1/m}$) and the function $y$
has a pole of order $j|n|$ at $(a_i,0)$, due to the pole structure of the differential $\varphi$, therefore $f(z)$ defined by \eqref{f2} and
considered as function on the $z$-sphere, has a pole of order $k$ at each point $z=a_i$.

Let us analyze the order of the zero of $f(z)$ at the point $z=\infty$. Consider the function $(J^*)^sy=(J^s)^*y$ evaluated at any point
$P_i\in X_a$ with $i=1,\dots, s$. On one hand, we know that $(J^*)^sy=\varepsilon^{jns} y$ and therefore
\begin{equation*}
(J^*)^sy(P_i) = \varepsilon^{jns}y(P_i)\,.
\end{equation*}
On the other hand, we have
\begin{equation*}
(J^*)^sy(P_i) = (J^s)^*y(P_i) = y(J^s(P_i))=y(P_i)\,.
\end{equation*}
The two above relations imply that $\varepsilon^{jns}y(P_i)=y(P_i)\,.$  Note that $y$ does not have a pole at $P_i$ as it would lead to a pole of $\varphi$ at $P_i\,.$  Therefore, given the assumption that $j$ is not a multiple of $m_1$,  we conclude that $jns$ is not a multiple of $m$ and thus $y(P_i)=0\,.$ The differential $dy = s\,\varphi$ vanishes at $P_i$ to the order $j|n|N_1-1$ (with respect to the local parameter $t=z^{-1/m_1}$ near this point). Thus we have that the function $y$ vanishes at any point $P_i$ with $i=1,\dots, s$ to the order $j|n|N_1$ and therefore the function $f=w^{-jn-km}y$ vanishes at $P_i$ to the order $kmN_1$.
Hence, as function on the $z$-sphere, $f(z)$ has a zero of order $kmN_1/m_1=kN$ at infinity.

Thus we obtain, similarly to the case $n>0$,
\begin{equation*}
f=w^{-jn-km}y = \frac{c}{ P^k(z)}
\end{equation*}
with some constant $c$ which may depend on the $\{a_i\}\,.$ From here we get $y=c\; P^{-k}(z)w^{jn+km}= c\;w^{jn}$ and thus
\begin{equation}
\label{concl2}
\varphi=\frac1s\,dy = c\; d w^{jn}\,.
\end{equation}

\item Finally,  let $n<0$ and suppose that $j$ is a multiple of $m_1\,,$ that is there is an integer $r$ such that $j=rm_1\,.$ Denote $h=(r,s)$ with $s=hs_1$ and $r=hr_1\,,$ where $r_1$ and $s_1$ are coprime. In this case, the surface $X_a$ can be seen as a ramified covering of the Riemann surface $\widehat X_a$ of the algebraic curve $\widehat w^{s_1} = P(z)$ with $\widehat w=w^{hm_1}\,.$
Differentials $\Omega_i^{(j)}(a)$ can be considered as being defined on $\widehat X_a:$
\begin{equation*}
\Omega_i^{(j)}(a)=\frac{w^{jn} dz}{(z-a_i)}
=\frac{w^{hr_1m_1n} dz}{(z-a_i)}
=\frac{\widehat w^{r_1n} dz}{(z-a_i)}
=:\widehat\Omega_i^{(r_1)}(a)
\end{equation*}
and differential $\varphi$ \eqref{phi} is also defined on $\widehat X_a$ as a linear combination of $\widehat\Omega_i^{(r_1)}(a)\;:$
\begin{equation*}
\varphi = \sum_{i=1}^{N-1}\alpha_i\Omega_i^{(j)}(a)
= \sum_{i=1}^{N-1}\alpha_i\widehat\Omega_i^{(r_1)}(a)\,.
\end{equation*}
Since $(r_1,s_1)=1\,,$ by the previous case of  $j$ being non divisible by  $m_1$ and a negative $n$ we have
\begin{equation}
\label{concl3}
\varphi=c\; d \widehat w^{r_1n}
=c\; d w^{jn}\,.
\end{equation}

\end{itemize}
Relations \eqref{concl1}, \eqref{concl2} and \eqref{concl3} imply
\begin{equation*}
\sum_{i=1}^{N-1}\alpha_i\frac{w^{jn} dz}{z-a_i} = c\; d w^{jn}\,.
\end{equation*}
Knowing that
\begin{equation*}
d w^{jn} = jn\; w^{jn}\,\frac{dw}w = \frac{jn}{m} w^{jn}\,\frac{dP}P= \frac{jn}{m} w^{jn}\sum_{i=1}^N\frac{dz}{z-a_i} \,,
\end{equation*}
the previous equality becomes
\begin{equation*}
\sum_{i=1}^{N-1}\alpha_i\frac{ dz}{z-a_i} = c\; \frac{jn}{m}  \sum_{i=1}^N\frac{dz}{z-a_i} \,.
\end{equation*}
Given that $\{a_i\}_{i=1}^N$ is an arbitrary set of distinct complex numbers, the above equality is only possible if $\alpha_i=c=0$ for all $i=1, \dots, N-1$ and thus the $N-1$ differentials $\Omega_1^{(j)},\ldots,\Omega_{N-1}^{(j)}$ are linearly independent in $H^1(X_a)\,.$ {\hfill $\Box$}

\section{Polynomial and rational solutions of  the  Schlesinger system}

Our differentials $\Omega_i^{(j)}(a)$  defined on the compact Riemann surface $X_a$  have poles at points at infinity
or at finite ramification points, depending on the sign of $n$. In general, the residues at these poles of $\Omega_i^{(j)}(a)$ are non-zero and, according to Theorem \ref{thm}, give rise to solutions of the Schlesinger system \eqref{hSchlesinger_intro}. In this section we show that such solutions are polynomial in $a_1,\ldots,a_N$ in the case of $n>0$ and rational in $a_1,\ldots,a_N$ in the case of $n<0$. This will lead us, in subsequent sections, to rational solutions of some Painlev\'e VI equations and to algebraic solutions of some Garnier systems.

\subsection{Polynomial solutions of  the  Schlesinger system}
\label{sect_thm2}

In this section we consider the case of $n>0$, when  differentials $\Omega_i^{(j)}(a)$ have poles at  $s$ points at infinity,  $s$ being the greatest common divisor of $m$ and $N$. Thus in the case of coprime $m$ and $N$ the residue of $\Omega_i^{(j)}$ at its only pole vanishes. In the case of $s>1$, however, $\Omega_i^{(j)}$ has $s$ poles with possibly non-zero residues,  which leads to the following statement on polynomial solutions of the Schlesinger system.

\begin{theorem}
\label{thm_2}  Let the eigenvalues of each matrix $B^{(i)}$, $i=1,\ldots,N$, have the same rational difference: $\beta_i^j-\beta_i^{j+1}=n/m$, $j=1,\ldots,p-1$, with $n>0$, $m>1$ coprime , and $s=(m,N)>1$ be the greatest common divisor of the integers $m$ and $N$. If there is an integer $j\in\{1,\ldots,p-1\}$
such that $sj/m\in{\mathbb Z}$, while $j/m\not\in{\mathbb Z}$, then the set of triangular solutions of system \eqref{hSchlesinger_intro} contains
a family of non-trivial polynomial ones:

$\bullet$ $b_i^{kl}(a)=c_{l-k}\,P_i^{(l-k)}(a)$, where $c_{l-k}\in\mathbb C$ is an arbitrary constant and $P_i^{(l-k)}$ is a non-zero polynomial of
degree $(l-k)\frac nm N$ given by \eqref{polynsol}, if $l$ and $k$ are such that $(l-k)s/m \in{\mathbb Z}$ and $(l-k)/m\not\in{\mathbb Z}$;

$\bullet$ $b_i^{kl}(a)\equiv0$ otherwise.
\end{theorem}

{\it Proof.} As explained in Section \ref{sect_local} for $n>0$, in the case $N=sN_1\,,$ $m=sm_1\,,$ where $N_1$, $m_1$ are coprime, each differential $\Omega_i^{(j)}(a)$
has $s$ poles $P_1,\ldots,P_s\in X_a$. In a local parameter $t$ at each pole $P_{\alpha}$ such that $t(P_{\alpha})=0$, the coordinate representation of
$\Omega_i^{(j)}$,  according to (\ref{omegab}), is of the form:
$$
\Omega_i^{(j)}=\frac{\nu_{\alpha}(1-a_1t^{m_1})^{jn/m}\ldots(1-a_Nt^{m_1})^{jn/m}}{t^{jnN_1+1}(1-a_it^{m_1})}\,dt,
\qquad\mbox{with}\quad \nu_{\alpha}=-m_1\,e^{2\pi\i jn(\alpha-1)/s}.
$$
Hence,
\begin{eqnarray}\label{omega_pole}
\Omega_i^{(j)}&=&\frac{\nu_{\alpha}\,dt}{t^{jnN_1+1}}\sum_{k_1=0}^\infty {jn/m\choose k_1} (-a_1t^{m_1})^{k_1}\ldots
\sum_{k_N=0}^\infty {jn/m\choose k_N}(-a_Nt^{m_1})^{k_N}\sum_{q=0}^\infty(a_it^{m_1})^q= \nonumber \\
         & = & \frac{\nu_{\alpha}\,dt}{t^{jnN_1+1}}\sum_{r=0}^\infty\Bigl[\sum_{k_1+\ldots+k_N+q=r}(-1)^{r-q}
				       {jn/m\choose k_1}\ldots{jn/m\choose k_N}a_1^{k_1}\ldots a_N^{k_N}a_i^q\Bigr]\,t^{rm_1},
\end{eqnarray}
where we use generalized binomial coefficients defined for any $\beta\in\mathbb R$ and $j\in\mathbb N$ by
\begin{equation*}
{\beta\choose j}=\frac{\beta(\beta-1)\cdots(\beta-j+1)}{j!}, \qquad {\beta\choose 0}=1.
\end{equation*}
Thus,  due to Theorem \ref{thm}, the integration of $\Omega_i^{(l-k)}(a)$, $i=1,\ldots,N$, along a small loop $\gamma_{l-k}$ encircling any pole $P_{\alpha}$ gives
$$
b_i^{kl}(a)=c_{l-k}\,\underset{P_{\alpha}}{\rm res}\,\Omega_i^{(l-k)}(a), \qquad c_{l-k}\in\mathbb C.
$$
As follows from (\ref{omega_pole}), the residue of $\Omega_i^{(l-k)}(a)$ equals zero if $(l-k)nN_1$ is not a multiple of $m_1\,,$ which is equivalent
to $l-k$ not being a multiple of $m_1$ because $m_1$ and $N_1$, as well as $m_1$ and $n$,  are coprime. Therefore, $b_i^{kl}(a)\equiv0$ if
$ (l-k)/m_1=(l-k)s/m\not\in\mathbb Z$.

In the case $(l-k)s/m$ is an integer,  denoting $d:=(l-k)\frac nm s$,  we have
\begin{equation}\label{polynsol}
\underset{P_{\alpha}}{\rm res}\,\Omega_i^{(l-k)}(a)=
\sum_{k_1+\ldots+k_N+q=N_1d}(-1)^q{d/s\choose k_1}\ldots{d/s\choose k_N}a_1^{k_1}\ldots a_N^{k_N}a_i^q
\end{equation}
up to an overall constant factor, that is $b_i^{kl}(a)$ is a polynomial of degree $N_1d=(l-k)\frac nm N$.  However, this polynomial is identically zero if
$(l-k)/m\in\mathbb Z$, since the differential $\Omega_i^{(l-k)}(a)$ is exact in this case. This finishes the proof of the theorem. {\hfill $\Box$}

\subsection{Rational solutions of  the  Schlesinger system}

In this section we consider the case of $n<0$, when the differentials $\Omega_i^{(j)}(a)$ have poles at the finite ramification points $(a_1,0),\ldots,(a_N,0)\in\hat\Gamma_a$. Contrary to the case of positive $n$, now the residues of $\Omega_i^{(j)}(a)$ at their poles are non-zero only if $j$ is a multiple of $m$ and we have the following statement on rational solutions of the Schlesinger system.

\begin{theorem}
\label{thm_ration}  Let the eigenvalues of each matrix $B^{(i)}$, $i=1,\ldots,N$, have the same rational difference: $\beta_i^j-\beta_i^{j+1}=n/m$, $j=1,\ldots,p-1$, with $n<0$, $m>0$ coprime. If there is an integer $j\in\{1,\ldots,p-1\}$
such that $j/m\in{\mathbb Z}$, then the set of triangular solutions of system \eqref{hSchlesinger_intro} contains
a family of non-trivial rational ones:

$\bullet$ $b_i^{kl}(a)=c_{l-k}\,R_i^{(l-k)}(a)$, if $l$ and $k$ are such that $(l-k)/m \in{\mathbb Z}$, where $c_{l-k}\in \mathbb C$ is an arbitrary constant and $R_i^{(l-k)}$ is a non-zero rational function given by \eqref{rationsol1}, for $i\ne\nu$, and by \eqref{rationsol2} for $i=\nu$, with an arbitrary number $\nu\in\{1,\ldots,N\}$ initially chosen;

$\bullet$ $b_i^{kl}(a)\equiv0$ otherwise.
\end{theorem}

{\it Proof.} We have the following parametrization of $\hat\Gamma_a$ near each ramification point $(a_{\nu},0)$ by a local parameter $t_\nu$:
$$
z=a_{\nu}+t_\nu^m, \qquad w=t_\nu\prod_{h=1,h\ne\nu}^N(a_{\nu}-a_h+t_\nu^m)^{1/m}, \qquad t_\nu\rightarrow0,
$$
whence the coordinate representation of $\Omega_i^{(j)}$ is of the form:
\begin{equation}\label{omegaram}
\Omega_i^{(j)}=\frac{w^{-j|n|}}{z-a_i}\,dz=\frac m{t_\nu^{j|n|-m+1}(a_{\nu}-a_i+t_\nu^m)}\prod_{h=1,h\ne\nu}^N(a_{\nu}-a_h+t_\nu^m)^{-j|n|/m}\,dt_\nu.
\end{equation}
Hence, for $i\ne\nu$ one has
\begin{eqnarray*}
\Omega_i^{(j)}&=&\frac m{t_\nu^{j|n|-m+1}}(a_{\nu}-a_i)^{-1}\Bigl(1+\frac{t_\nu^m}{a_{\nu}-a_i}\Bigr)^{-1}\prod_{h=1,h\ne\nu}^N(a_{\nu}-a_h)^{-j|n|/m}\Bigl(1+\frac{t_\nu^m}{a_{\nu}-a_h}\Bigr)^{-j|n|/m}\,dt_\nu=\\
&=& \frac{m\,dt_\nu}{t_\nu^{j|n|-m+1}}\sum_{r=0}^\infty\Bigl[\sum_{k_1+\ldots+k_N=r}\frac{(-1)^{k_{\nu}}}{(a_{\nu}-a_i)^{k_{\nu}+1}}\prod_{h=1,h\ne\nu}^N\frac{{-j|n|/m\choose k_h}}{(a_{\nu}-a_h)^{k_h+j|n|/m}}\Bigr]\,t_\nu^{rm},
\end{eqnarray*}
while
\begin{eqnarray*}
\Omega_{\nu}^{(j)}&=&\frac m{t_\nu^{j|n|+1}}\prod_{h=1,h\ne\nu}^N(a_{\nu}-a_h)^{-j|n|/m}\Bigl(1+\frac{t_\nu^m}{a_{\nu}-a_h}\Bigr)^{-j|n| /m}\,dt_\nu=\\
&=& \frac{m\,dt_\nu}{t_\nu^{j|n|+1}}\sum_{r=0}^\infty\Bigl[\sum^{\prime}_{k_1+\ldots+k_N=r}\prod_{h=1,h\ne\nu}^N\frac{{-j|n|/m \choose k_h}}{(a_{\nu}-a_h)^{k_h+j|n|/m}}\Bigr]\,t_\nu^{rm},
\end{eqnarray*}
where the summation index $k_{\nu}$ is missed in the above sum $\sum^{\prime}$.

Like in the previous theorem, the integration of $\Omega_i^{(l-k)}(a)$, $i=1,\ldots,N$, along a small loop $\gamma_{l-k}$ encircling any pole $(a_{\nu},0)$ gives
$$
b_i^{kl}(a)=c_{l-k}\,\underset{(a_{\nu},0)}{\rm res}\,\Omega_i^{(l-k)}(a), \qquad c_{l-k}\in\mathbb C.
$$
As follows from the above coordinate representation, the residue of $\Omega_i^{(l-k)}(a)$ equals zero if $(l-k)n$ is not a multiple of $m\,,$ which is equivalent to $l-k$ not being a multiple of $m$. Therefore, $b_i^{kl}(a)\equiv0$ if
$ (l-k)/m\not\in\mathbb Z$.

In the case $(l-k)/m$ is an integer,  denoting $d:=(l-k)|n|/m$,  we have
\begin{equation}\label{rationsol1}
\underset{(a_{\nu},0)}{\rm res}\,\Omega_i^{(l-k)}(a)=\sum_{k_1+\ldots+k_N=d-1}\frac{(-1)^{k_{\nu}}}{(a_{\nu}-a_i)^{k_{\nu}+1}}\prod_{h=1,h\ne\nu}^N{-d\choose k_h}\frac1{(a_{\nu}-a_h)^{k_h+d}}, \quad i=1,\ldots,N, \; i\ne\nu,
\end{equation}
up to an overall constant factor, and
\begin{equation}\label{rationsol2}
\underset{(a_{\nu},0)}{\rm res}\,\Omega_{\nu}^{(l-k)}(a)=\sum^{\prime}_{k_1+\ldots+k_N=d}\prod_{h=1,h\ne\nu}^N{-d\choose k_h}\frac1{(a_{\nu}-a_h)^{k_h+d}}.
\end{equation}
This finishes the proof of the theorem. {\hfill $\Box$}

\section{Application to Painlev\'e VI equations}\label{Painleve}

As is well known, in the case $p=2$, $N=3$ (assuming $(a_1,a_2,a_3)=(0,1,x)$, $x\in{\mathbb C}\setminus\{0,1\}$) the Schlesinger system
for {\it traceless} $(2\times2)$-matrices $B^{(1)}(x)$, $B^{(2)}(x)$, $B^{(3)}(x)$,
\begin{equation}\label{Schl22}
\frac{dB^{(1)}}{dx} = \frac{[B^{(3)}, B^{(1)}]}{x}\,, \qquad
\frac{dB^{(2)}}{dx} = \frac{[B^{(3)}, B^{(2)}]}{x-1}\,, \qquad
B^{(1)} +B^{(2)}+B^{(3)} = \left(\begin{array}{cc}-\beta_{\infty} & 0 \\0 & \beta_{\infty}\end{array}\right)
\end{equation}
 (if $\beta_{\infty}=0$, the last matrix sum is a Jordan cell),  corresponds to the sixth Painlev\'e equation ${\rm P_{VI}}(\alpha,\beta,\gamma,\delta)$

\begin{equation*}
\frac{d^2 y}{dx^2}\!\! =\!\! \frac{1}{2}\!\! \left(  \frac{1}{y} + \frac{1}{y-1}+ \frac{1}{y-x}  \right) \!\!\left( \frac{dy}{dx} \right)^2 \!\!- \left( \frac{1}{x} + \frac{1}{x-1} + \frac{1}{y-x} \right) \frac{dy}{dx}
\end{equation*}
\begin{equation*}
 + \frac{y(y-1)(y-x)}{x^2(x-1)^2}\left( \alpha +\beta\frac{x}{y^2} + \gamma\frac{x-1}{(y-1)^2} + \delta\frac{x(x-1)}{(y-x)^2}  \right).
\end{equation*}

The parameters $(\alpha,\beta,\gamma,\delta)$ of ${\rm P_{VI}}$  are computed from the
eigenvalues $\pm\beta_i$ of the matrices $B^{(i)}$, $i=1,2,3$, as follows:
$$
\alpha = \frac{(2\beta_\infty-1)^2}{2}, \qquad \beta=-2\beta_1^2, \qquad \gamma=2\beta_2^2, \qquad \delta = \frac{1}{2} - 2\beta_3^2.
$$
Namely, the function
\begin{equation}\label{P6sol}
y(x)=\frac{xb_1}{b_1+(1-x)b_3}\,,
\end{equation}
where $b_i$ is a $(1,2)$-entry of the matrix $B^{(i)}$, satisfies the Painlev\'e VI with the above parameters.

In our triangular case, solutions
\begin{equation}\label{sol22}
B^{(i)} = \left(\begin{array}{cc}\beta_i & b_i(x) \\0 & -\beta_i\end{array}\right)\,, \qquad i=1,2,3,
\end{equation}
of the Schlesinger system (\ref{Schl22}) are hypergeometric. For example, as a consequence of the Schlesinger equations, the functions
$b_1$ and $b_2$ satisfy the following linear differential system:
\begin{eqnarray}
\label{systembb}
\left\{\begin{array}{l}b_1'=\frac{2}{x}\left( (\beta_1+\beta_3) b_1 + \beta_1 b_2 \right) \\
\\
b_2'=\frac{2}{x-1}\left(\beta_2 b_1+ (\beta_2+\beta_3)b_2  \right)
\end{array}\right.
\end{eqnarray}
and thus solve the hypergeometric linear differential equations of the form (see  \cite[Ch. 4, \S3.3]{FIKN})
\begin{equation}
\label{hypergeom1}
b_1''+\frac{(2\beta_1+2\beta_3-1)+(1-2\beta_1-2\beta_2-4\beta_3)x}{x(x-1)}\,b_1'+
\frac{4\beta_3(\beta_1+\beta_2+\beta_3)}{x(x-1)}\,b_1=0\,,
\end{equation}
\begin{equation}
\label{hypergeom2}
b_2''+\frac{(2\beta_1+2\beta_3)+(1-2\beta_1-2\beta_2-4\beta_3)x}{x(x-1)}\,b_2'+
\frac{4\beta_3(\beta_1+\beta_2+\beta_3)}{x(x-1)}\,b_2=0\,,
\end{equation}
while $b_3=-b_1-b_2\,.$

This means that solutions of a {\it triangular} Schlesinger system (\ref{Schl22}) always lead to hypergeometric solutions
of the corresponding sixth Painlev\'e equation through \eqref{P6sol}. More precisely, from a general two-parameter family
of solutions of \eqref{hypergeom1}  linearly  parameterized by constants $c_1$, $c_2$, one obtains $b_2$
using the first equation of \eqref{systembb}, and then $b_3=-b_1-b_2$. A particular one-parameter family of solutions of the
corresponding sixth Painlev\'e equation parametrized by the ratio $c_1/c_2$ is then obtained by \eqref{P6sol}.

In the case we consider, the eigenvalues in (\ref{sol22}) are given by
$$
\beta_1=\beta_2=\beta_3=\frac{n}{2m}, \qquad \mbox{and} \quad \beta_\infty = -\frac{3n}{2m},
$$
with any coprime integers $n>0$, $m>1$  or $n<0$, $m>0$. Applying Theorems \ref{thm}  and \ref{thm_conj}  we obtain algebro-geometric expressions for a  one-parameter family of hypergeometric solutions $y(x)$ of the sixth Painlev\'e equation
${\rm P_{VI}}\Bigl(\frac{(3n+m)^2}{2m^2},-\frac{n^2}{2m^2},\frac{n^2}{2m^2},\frac{m^2-n^2}{2m^2}\Bigr)$:
\begin{equation}\label{b22}
y(x)=\frac{xb_1}{b_1+(1-x)b_3},
\end{equation}
\begin{equation*}
 b_1 =\oint_{\gamma_1}\frac{w^{n}dz}z+c\oint_{\gamma_2}\frac{w^{n}dz}z\,,\qquad
 b_3 =\oint_{\gamma_1}\frac{w^{n}dz}{z-x}+c\oint_{\gamma_2}\frac{w^{n}dz}{z-x}\,, \qquad  c\in{\mathbb C},
\end{equation*}
where $\gamma_1,\gamma_2$ are suitable closed contours on the Riemann surface $X_x$ of the curve
$$
w^m=z(z-1)(z-x)
$$
with the only variable branch point $x\in\mathbb C\setminus\{0,1\}$ (or, on the $X_x$ punctured at three points, the poles of the differentials $w^ndz/z$, $w^ndz/(z-x)$, $w^ndz/(z-1)$,  depending on which of the cases (a), (b), (c) of Theorem \ref{thm} holds).

\subsection{Rational solutions of $P_{\rm VI}$: a torus with three punctures}
\label{sect_rationalPVI_torus}

In this section we consider the case (b) of Theorem \ref{thm} in the context of Painlev\'e VI equations, that is, the case
of $n>0$, $m>1$, $p=2$ and $s=(m,N)=(m,3)=3$. Let us analyze the requirements of Theorem \ref{thm_2} in this case and
see when we can apply this theorem to obtain polynomial expressions for the $b_i$'s.

As $s=3$, the requirement $s/m\in{\mathbb Z}$ of Theorem \ref{thm_2} implies that $m=3$. Hence we deal with the Riemann surface $X_x$ of the curve
$$
w^3=z(z-1)(z-x)
$$
punctured at three points $P_1, P_2, P_3$ at infinity. The genus of $X_x$ equals
$$
g=\frac12\bigl((m-1)(N-1)-s+1\bigr)=1,
$$
that is, this is a torus and there are four basic cycles on $X_x\setminus\{P_1,P_2,P_3\}$.

Computing the residues $b_1(x)$, $b_2(x)$, $b_3(x)$ of the differentials $w^ndz/z$, $w^ndz/(z-1)$, $w^ndz/(z-x)$, say at the pole $P_1$, 
we obtain polynomial solutions (\ref{sol22}) of the Schlesinger system (\ref{Schl22}), with $\beta_1=\beta_2=\beta_3=n/6\not\in\frac12\mathbb Z$ and $\beta_{\infty}=-n/2$.  Namely, the coordinate representation of the above differentials in a local parameter $t$ such that $t(P_1)=0$,
according to formula (\ref{omegab}) with $N=3$, $(a_1,a_2,a_3)=(0,1,x)$ and $m=3$, $m_1=N_1=1$, is of the form
$$
\frac{w^ndz}z=-(1-t)^{n/3}\,(1-xt)^{n/3}\,\frac{dt}{t^{n+1}}, \qquad \frac{w^ndz}{z-1}=-(1-t)^{n/3-1}\,(1-xt)^{n/3}\,\frac{dt}{t^{n+1}},
$$
$$
\frac{w^ndz}{z-x}=-(1-t)^{n/3}\,(1-xt)^{n/3-1}\,\frac{dt}{t^{n+1}}.
$$
The first differential has therefore the following expansion near $t=0$:
$$
-\sum_{j_1=0}^{\infty}{n/3\choose j_1}(-t)^{j_1}\sum_{j_2=0}^{\infty}{n/3\choose j_2}(-xt)^{j_2}\,\frac{dt}{t^{n+1}},
$$
whence its residue $b_1(x)$ at $t=0$ equals, up to a constant factor of $(-1)^{n+1}$,
\begin{eqnarray*}
\label{b1_poly}
b_1(x)=b_1^{\rm P}(x)=\sum_{j=0}^n{n/3\choose j}{n/3\choose n-j} x^j\,. \label{b1}
\end{eqnarray*}
Similarly, for the residues $b_2(x)$, $b_3(x)$ of the other two differentials, up to the same factor of $(-1)^{n+1}$, one has
\begin{eqnarray*}
\label{b2_poly}
b_2(x)=b_2^{\rm P}(x)&=&\sum_{j=0}^n{n/3\choose j}{n/3-1\choose n-j}x^j\,,\;\; \label{b2} \\
\label{b3_poly}
b_3(x)=b_3^{\rm P}(x)&=&\sum_{j=0}^n{n/3-1\choose j}{n/3\choose n-j}x^j\,.\label{b3}
\end{eqnarray*}
The functions $b_1^{\rm P}$ and $b_2^{\rm P}$ are related to each other by system \eqref{systembb}. They give degree $n$ polynomial solutions to
the hypergeometric equations \eqref{hypergeom1} and \eqref{hypergeom2}, respectively. Furthermore, the polynomials $xb_1^{\rm P}(x)$ and
$$
b_1^{\rm P}(x)+(1-x)b_3^{\rm P}(x)=-b_2^{\rm P}(x)-xb_3^{\rm P}(x)=-\,\frac{3(n+1)}n\sum_{j=0}^{n+1}{n/3\choose j}{n/3\choose n+1-j} x^j
$$
give, {\it via} \eqref{b22}, a rational solution to the Painlev\'e VI equation with the parameters
$$
(\alpha,\beta,\gamma,\delta)=\left(\frac{(2\beta_\infty-1)^2}2, -2\beta_1^2, \, 2\beta_2^2, \, \frac12-2\beta_3^2\right)=
                             \left(\frac{(n+1)^2}2, -\frac{n^2}{18}, \, \frac{n^2}{18}, \, \frac{9-n^2}{18}\right),
$$
and thus we obtain the following assertion.
\medskip

\begin{theorem}\label{thmrat1}
For every positive integer $n$ not divisible by $3$, the polynomials
$$
 P_{n+1}(x)=x\,\sum_{j=0}^n{n/3\choose j}{n/3\choose n-j} x^j  \quad\mbox{and}\quad
Q_{n+1}(x)=-\frac{3(n+1)}n\sum_{j=0}^{n+1}{n/3\choose j}{n/3\choose n+1-j} x^j
$$
of degree $n+1$ define the rational solution $y_n(x)=P_{n+1}(x)/Q_{n+1}(x)$
of the sixth Painlev\'e equation ${\rm P_{VI}}(\alpha,\beta,\gamma,\delta)$ with the parameters
$$
\alpha=\frac{(n+1)^2}2, \quad \beta=-\frac{n^2}{18}, \quad \gamma=\frac{n^2}{18}, \quad \delta=\frac{9-n^2}{18}.
$$
\end{theorem}

Note that none of the monodromy matrices $M_1,M_2,M_3$ of the triangular Schlesinger isomonodromic family corresponding to the above
$b_i^{\rm P}$'s, at the points $z=0$, $z=1$, $z=x$ respectively, equals $\pm\mathbb I$, since the eigenvalues
$e^{\pm2\pi{\bf i}\beta_i}=e^{\pm\pi{\bf i}n/3}$ of each $M_i$ do not equal $\pm1$. Therefore, due to Lemma 3.3 from \cite{Maz11}, the monodromy of
this family is commutative. In fact, the commutativity of the monodromy of a Schlesinger isomonodromic family is a general necessary condition for
the corresponding solution of the sixth Painlev\'e equation to be rational, see Remark \ref{rmkRat} below.

\begin{example}
\label{example_1}
{\rm Let us compute degree $n$ polynomial solutions to the hypergeometric equations \eqref{hypergeom1}, \eqref{hypergeom2}, with
$\beta_1=\beta_2=\beta_3=n/6$, and a rational solution to the corresponding Painlev\'e VI equation in the case $n=1$, $n=2$, and $n=4$.
\begin{enumerate}
\item For $n=1$, we obtain the following linear functions:
\begin{equation*}
b_1^{\rm P}(x)=\frac{x+1}{3}\;;
\qquad
b_2^{\rm P}(x)=\frac{x-2}{3}\;;
\qquad
b_3^{\rm P}(x)=\frac{-2x+1}{3}\;,
\end{equation*}
where $b_1^{\rm P}$ satisfies \eqref{hypergeom1} and $b_2^{\rm P}$ satisfies \eqref{hypergeom2}, with $\beta_1=\beta_2=\beta_3=1/6$. The corresponding
rational solution of the sixth Painlev\'e equation ${\rm P_{VI}}\bigl(2, -\frac{1}{18}, \frac{1}{18}, \frac{4}{9}\bigr)\,$ is given by
\begin{equation*}
y_1(x)= \frac{x(x+1)}{2x^2-2x+2}\;.
\end{equation*}
\item For $n=2$, we obtain the functions
\begin{equation*}
b_1^{\rm P}(x)=-\frac{1}{9}\left( x^2-4x+1 \right)\;;
\qquad
b_2^{\rm P}(x)=-\frac{1}{9}\left(  x^2+2x-2\right)\;;
\qquad
b_3^{\rm P}(x)=-\frac{1}{9}\left( -2x^2+2x+1 \right)\;,
\end{equation*}
leading to the following rational solution of the sixth Painlev\'e equation
${\rm P_{VI}}\bigl(\frac{9}{2}, -\frac{2}{9}, \frac{2}{9}, \frac{5}{18}\bigr)\,:$
\begin{equation*}
y_2(x)= \frac{x(x^2-4x+1)}{2x^3-3x^2-3x+2}\,.
\end{equation*}
\item For $n=4$, we get the functions
\begin{equation*}
b_1^{\rm P}(x)=\frac{1}{3^5}\left( 5x^4-16x^3+12x^2-16x+5  \right)\;;
\qquad
b_2^{\rm P}(x)=\frac{1}{3^5}\left( 5x^4-4x^3-6x^2+20x-10 \right)\;;
\end{equation*}
\begin{equation*}
b_3^{\rm P}(x)=\frac{1}{3^5}\left( -10x^4+20x^3-6x^2-4x+5 \right)\;,
\end{equation*}
leading to the following rational solution of the sixth Painlev\'e equation
${\rm P_{VI}}\bigl(\frac{25}{2}, -\frac{8}{9}, \frac{8}{9}, -\frac{7}{18}\bigr)\,:$
\begin{equation*}
y_4(x)= \frac{x(5 x^4 - 16 x^3 + 12 x^2 - 16 x + 5)}{10 x^5 - 25 x^4 + 10 x^3 +
   10 x^2 - 25 x + 10}\,.
\end{equation*}
\end{enumerate}
}
\end{example}

\begin{figure}[h]
\centering
{\includegraphics[width=7cm]{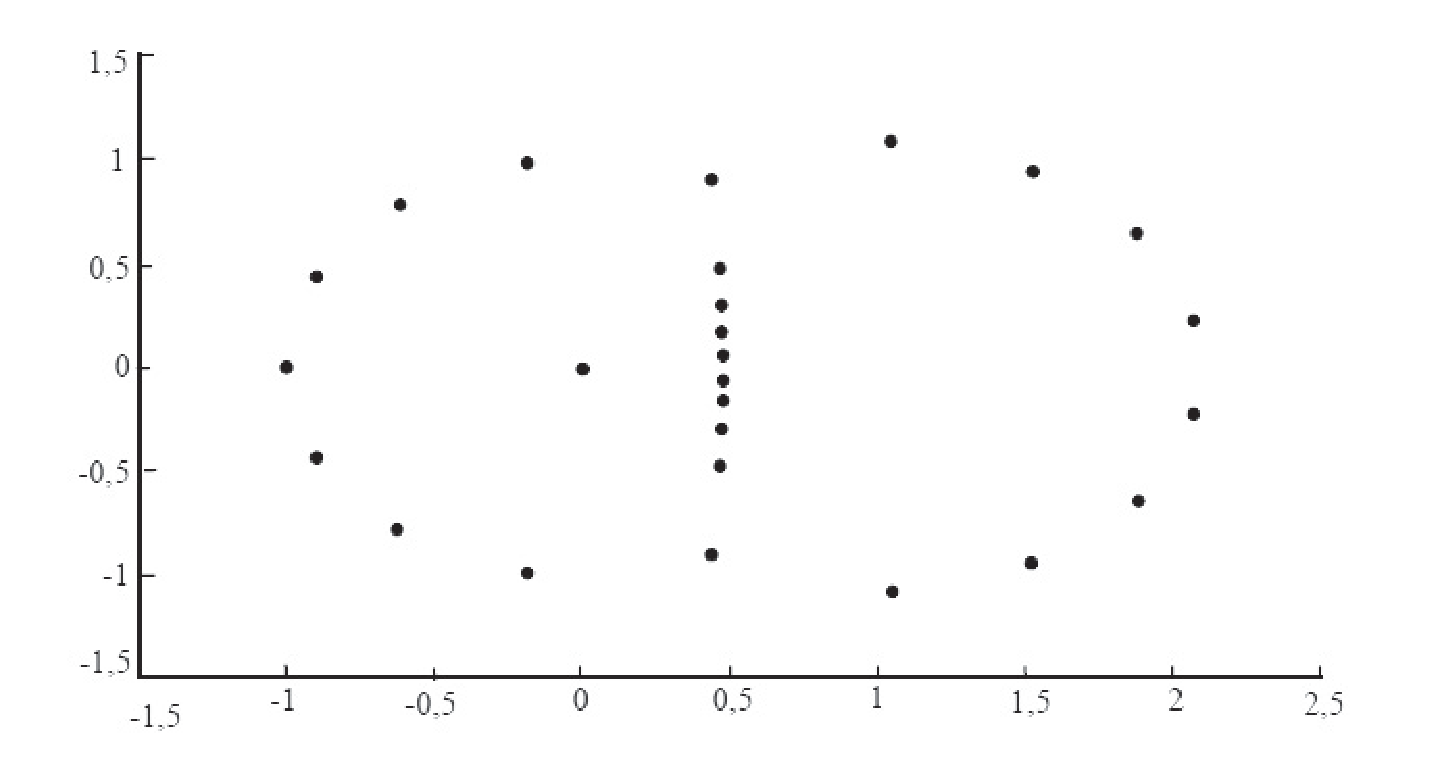}
\includegraphics[width=7cm]{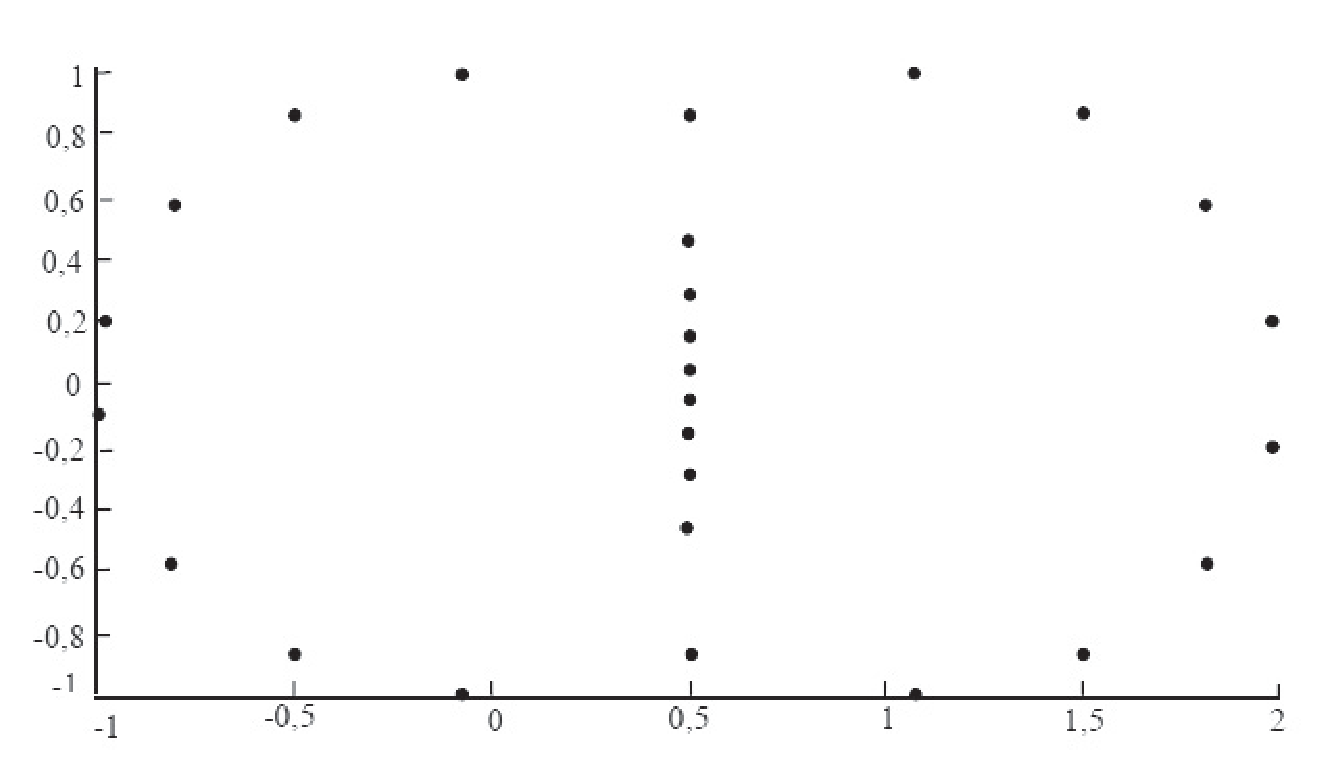}}
\caption{Distribution of zeros for $P_{26}$ and $Q_{26}$.}\label{fig:zeros26}
\end{figure}

\begin{figure}[h]
\centering
{\includegraphics[width=7cm]{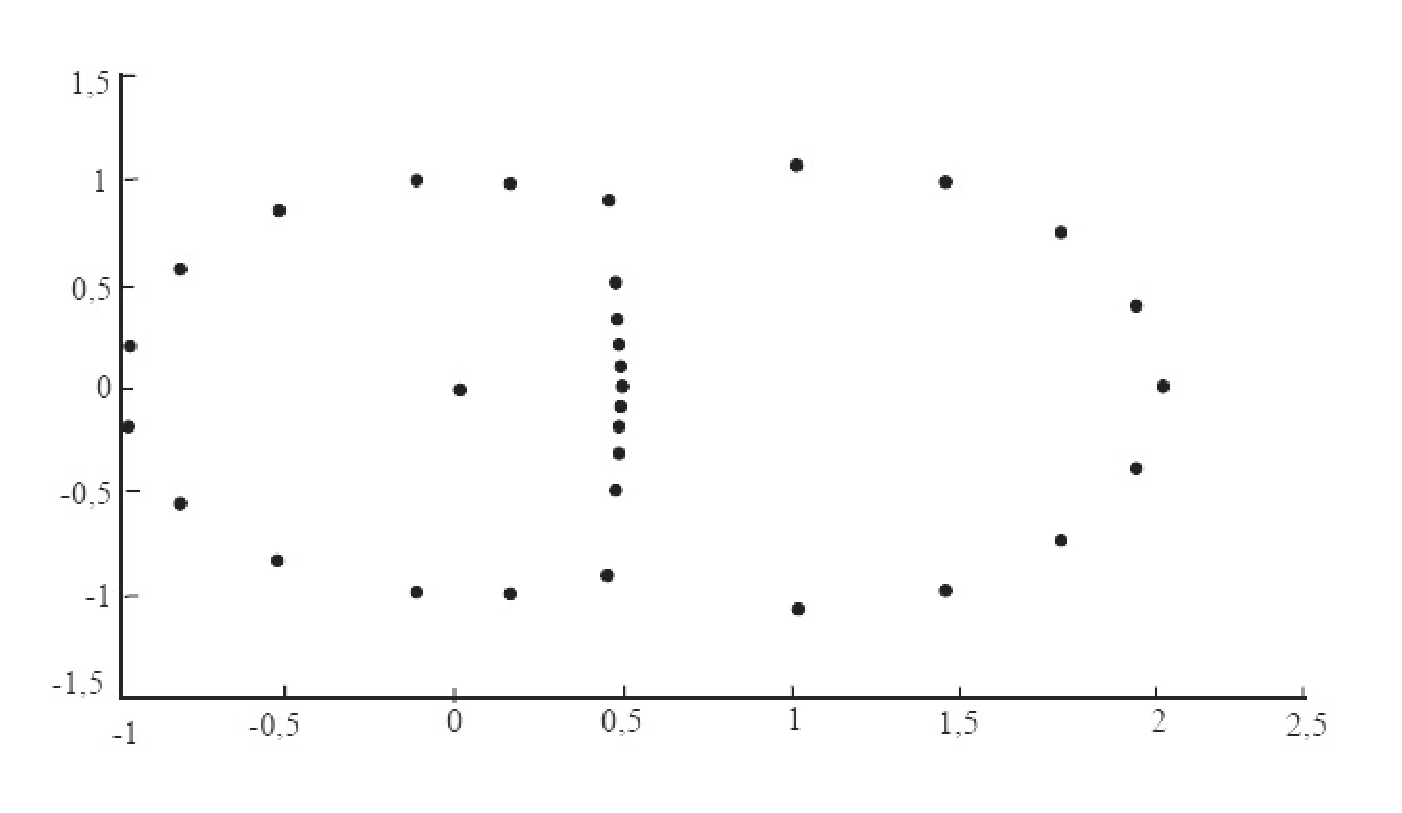}
\includegraphics[width=7cm]{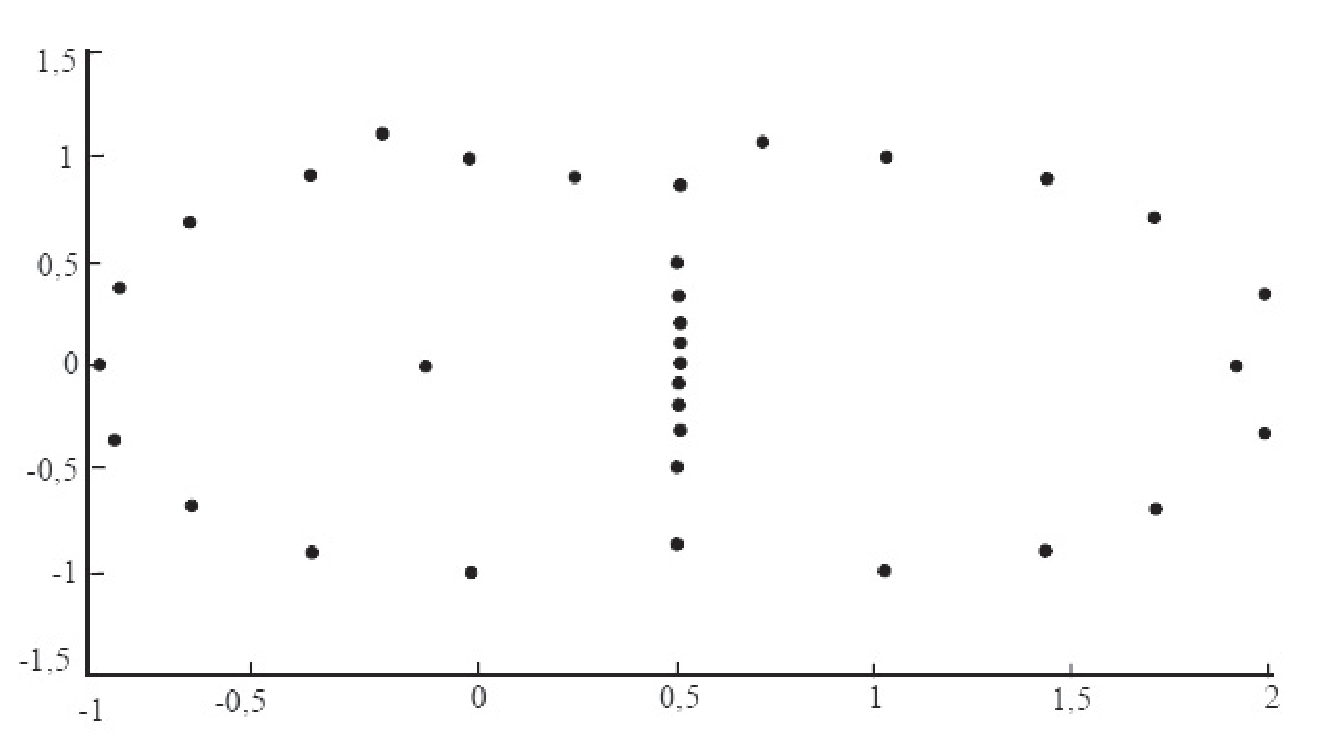}}
\caption{Distribution of zeros for $P_{29}$ and $Q_{29}$.}\label{fig:zeros29}
\end{figure}

The polynomials  $b_1^{\rm P}(x)=P_{n+1}(x)/x$  and $ Q_{n+1}(x)$ are both reciprocal. Let us recall that a polynomial
of degree $n$ of the form $\sum_{j=0}^na_jx^j$ is {\it reciprocal} if $a_j=a_{n-j}$, for all $j=0, \dots, n$. Thus, for $n=2k+1$ odd, the polynomial $P_{n+1}(x)$ has zeros at $0, -1$ and $k$ pairs of zeros $z_j, z_j^{-1}$, while the polynomial $Q_{n+1}(x)$ has $k+1$ pairs of zeros $w_j, w_j^{-1}$.
For $n=2k$ even, the polynomial $P_{n+1}(x)$ has a zero at $0$ and $k$ pairs of zeros $z_j, z_j^{-1}$, while the polynomial $Q_{n+1}(x)$ has a zero at
$-1$ and $k$ pairs of zeros $w_j, w_j^{-1}$. Since the polynomials have all coefficients real, this implies that the roots of each polynomial are situated symmetrically with respect to the real axis  and all roots different from $0$ are placed symmetrically  with respect to the unit circle in the sense of inversion. Figures \ref{fig:zeros26} and \ref{fig:zeros29} show the distribution of zeros of $P$ and $Q$ with $n=25$ and $n=28$.
 These intriguing patterns are explained by results of A.\,Kuijlaars, A.\,Martinez-Filkenshtein \cite{KMF} as was pointed out to us by a referee; we detail this now.

Let us recall that the hypergeometric equation
\begin{equation}\label{hypergeomgen}
x(1-x)\,b_1''+[ {\bf c}-({\bf a}+{\bf b}+1)  x]\,b_1'- {\bf ab}\,b_1=0,
\end{equation}
possesses a solution in the form of a hypergeometric series:
$$
F({\bf a},{\bf b, c}, x)=\sum_{j=0}^{\infty}\frac{({\bf a})_j\,({\bf b})_j}{({\bf c})_j\,j!}\,x^j,
$$
where $(\theta)_j=\theta(\theta+1)\ldots(\theta+j-1)$, $(\theta)_0=1$, for any $\theta\in\mathbb C$.
For ${\bf a}=-n$, with  $n\in\mathbb N$, the above series truncates and we get polynomials:
$$
F(-n,{\bf b, c}, x)=\sum_{j=0}^{\infty}\frac{(-n)_j\,({\bf b})_j}{({\bf c})_j\,j!}\,x^j=
\sum_{j=0}^n\frac{(-n)_j\,({\bf b})_j}{({\bf c})_j\,j!}\,x^j.
$$
It is easy to check that the polynomials $P_{n+1}$, $Q_{n+1}$ obtained in Theorem \ref{thmrat1} are given by
$$
P_{n+1}(x)={n/3\choose n} x\,F(-n,-n/3,1-2n/3,x), \qquad Q_{n+1}(x)=2{n/3\choose n} F(-n-1,-n/3,-2n/3,x),
$$
which agrees with the fact that the polynomial  $b_1^{\rm P}(x)=P_{n+1}(x)/x$  is a solution of the hypergeometric
equation \eqref{hypergeomgen} with
\begin{equation}\label{hyperparam}
{\bf a}=-2(\beta_1+\beta_2+\beta_3)=-n, \qquad {\bf b}=-2\beta_3=-n/3, \qquad {\bf c}=1-2(\beta_1+\beta_3)=1-2n/3
\end{equation}
(that is, a solution of (\ref{hypergeom1}) with $\beta_1=\beta_2=\beta_3=n/6>0$ and $\not\in\frac12\mathbb Z$).

On the other hand, Jacobi polynomials $P_n^{(\alpha, \beta)}(t)$ can be defined as (see e.g. \cite[Sect. 6.3]{AAR}, \cite[Sect. 4.21]{Sz}):
\begin{equation}\label{eq:JacobiPol}
P_n^{(\alpha, \beta)}(t)={n+\alpha\choose n}F\bigl(-n, n+\alpha +\beta+1, \alpha+1, \frac{1-t}{2}\bigr).
\end{equation}
Thus we get

\begin{lemma}\label{lemma:Jacobi}
The polynomials $P_{n+1}(x)$ and $Q_{n+1}(x)$ from Theorem \ref{thmrat1} can be expressed through Jacobi polynomials \eqref{eq:JacobiPol}
as follows:
$$
P_{n+1}(x) = x\,P_n^{(\alpha_n, \beta_n)}(1-2x), \qquad Q_{n+1}(x) =-\frac{3(n+1)}n\,P_{n+1}^{(\hat\alpha_{n+1}, \hat\beta_{n+1})}(1-2x),
$$
where
\begin{equation}\label{eq:alphabeta}
\alpha_n=-\frac{2n}{3}, \quad \beta_n=-\frac{2n}{3}-1; \qquad \hat\alpha_n=-\frac{2n+1}{3}, \quad
\hat\beta_n=-\frac{2n+1}{3}.
\end{equation}
\end{lemma}

From \eqref{eq:alphabeta} we have
$$
A=\lim_{n\rightarrow\infty}\frac{\alpha_n}{n}=\lim_{n\rightarrow\infty}\frac{\hat\alpha_n}{n}=-\frac23, \qquad B=\lim_{n\rightarrow\infty}\frac{\beta_n}{n}=\lim_{n\rightarrow\infty}\frac{\hat\beta_n}{n}=-\frac23.
$$
The study of asymptotics of zeros of Jacobi polynomials $P_n^{(\alpha_n, \beta_n)}$ with
$-1<A<0$, $-1<B<0$, and $-2<A+B<-1$ was done in \cite{KMF}. More precisely, one can apply Theorem 2.3 from \cite{KMF} and get the following

\begin{figure}[h]
\centering
{\includegraphics[width=10cm]{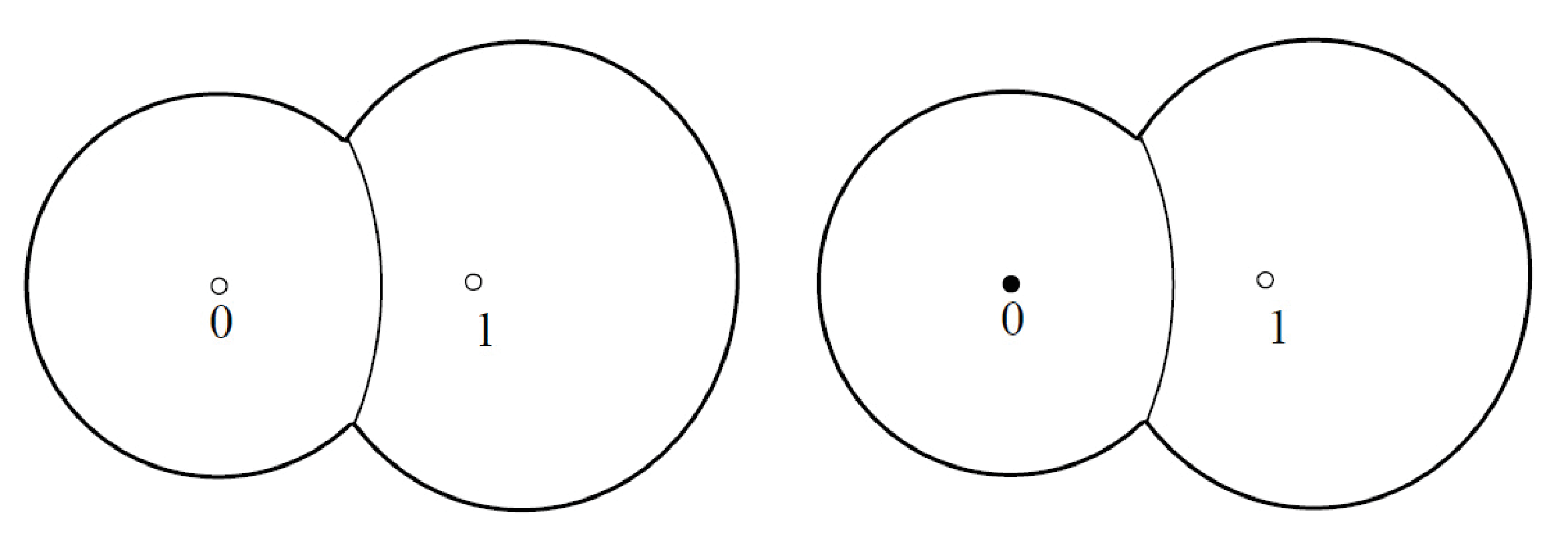}}
\caption{Contours $\Gamma_{A,B}$  and $\Gamma_{A,B}\cup \{0\}$ where the zeros of $Q_{n+1}$ and $P_{n+1}$ accumulate in the limit $n\rightarrow \infty $.}\label{fig:contour1}
\end{figure}

\begin{proposition}\label{th:zeros} As $n\rightarrow \infty$ the zeros of polynomials $Q_{n+1}(x)$ accumulate on a contour $\Gamma_{A,B}$, with $A=B=-2/3$, as presented in Fig. \ref{fig:contour1} left, while the zeros of polynomials $P_{n+1}(x)$ accumulate on the same contour $\Gamma_{A,B}$ with the point $\{0\}$
added to it, see Fig. \ref{fig:contour1} right.
\end{proposition}

An explicit analytical description of the contour $\Gamma_{A,B}$ follows from Section 2.1 of \cite{KMF}. In particular, it is formed by three
analytic arcs intersecting at two points $\zeta_{\pm}=1/2\pm{\rm i}\sqrt3/2$. The leftmost arc is a part of the unit circle centered at the origin, the central and right arcs are placed symmetrically with respect to this circle in the sense of inversion. One can compare Figures \ref{fig:zeros26}, \ref{fig:zeros29}, and \ref{fig:contour1} with Figures 2 and 3 from \cite{KMF} noting that our contours are related to those from \cite{KMF} by an affine transformation $1-2x$.

It is known that rational solutions of Painlev\'e equations can typically be expressed in terms of logarithmic derivatives of special polynomials that are defined through second order recursion relations, and that these solutions possess a determinant structure and their zeros have a highly symmetric and regular behaviour. For the Painlev\'e equations II--V, see P.\,Clarkson's expositions \cite{Clark1}, \cite{Clark2} explaining these issues and references therein.
We would mention G.\,Almqvist's contribution \cite{Alm3} as an example of research done in this direction for Painlev\'e VI. Proposition
\ref{th:zeros} above and the following proposition indicate the possibilities of including our rational solutions in this context.

\begin{proposition}\label{prop:rat1} The solutions $y_n(x)$ from Theorem \ref{thmrat1}
can be rewritten in terms of the Jacobi polynomials \eqref{eq:JacobiPol} with $\alpha_n=-2n/3$ as follows:
\begin{eqnarray}\label{eq:PailneveVIJacobi1}
{\rm (a)}\qquad y_n(x)&=& \frac{1}{2} - \frac{n}{6(n+1)}\,\frac{P_n^{(\alpha_n-1,\alpha_n-1)}(1-2x)}{P_{n+1}^{(\alpha_n-1,\alpha_n-1)}(1-2x)}; \\
{\rm (b)}\qquad y_n(x)&=& \frac{1}{2} - \frac{n}{2(n+1)(n+6)}\,\frac{\frac{d}{dx}P_{n+1}^{(\alpha_n-2,\alpha_n-2)}(1-2x)}{P_{n+1}^{(\alpha_n-1,\alpha_n-1)}(1-2x)}.
\label{eq:PailneveVIJacobi2}
\end{eqnarray}
\end{proposition}

{\it Proof.} The proof follows from Theorem \ref{thmrat1}, Lemma \ref{lemma:Jacobi}, and the following known identities for Jacobi polynomials
(see e.g. \cite[Sect. 6.4]{AAR}, \cite[Sect. 4.5]{Sz}):
\begin{eqnarray*}
(n+\alpha+1)P_n^{(\alpha, \beta)}(x)-(n+1)P_{n+1}^{(\alpha, \beta)}(x)&=&\frac{2n+\alpha +\beta +2}{2}(1-x)P_n^{(\alpha+1, \beta)}(x); \\
\frac{d}{dx} P_n^{(\alpha, \beta)}(x)&=&\frac{n+\alpha+\beta+1}{2}\,P_{n-1}^{(\alpha+1, \beta+1)}(x).
\end{eqnarray*}
{\hfill $\Box$}

\begin{remark}\label{rmkRat}
{\rm All Painlev\'e VI equations which have {\it non-degenerate}  rational solutions were classified by  M.\,Mazzocco
\cite{Maz1}. We will refer to a more recent arXiv version \cite{Maz11}, where some instances of \cite{Maz1} were formulated differently.
M.\,Mazzocco proved that they occur if and only if for the corresponding Schlesinger system $(\ref{Schl22})$ there holds
$$
\beta_{\infty}+\varepsilon_1\beta_1+\varepsilon_2\beta_2+\varepsilon_3\beta_3\in{\mathbb Z},
$$
for some choice of $\varepsilon_i\in\{\pm1\}$ and at least one $\beta_i\in\frac12\mathbb Z$.  The monodromy of the corresponding Schlesinger isomonodromic family is necessarily {\it commutative}. As stated in \cite{Maz11}, all such  rational solutions are equivalent, {\it via}
Okamoto's  birational canonical transformations \cite{Ok2} and up to symmetries, to the following solutions:
\begin{eqnarray}
y(x)=\frac x{(1+2\beta_3)+(1+2\beta_2)x}\;, & \quad & \beta_{\infty}+\beta_1+\beta_2+\beta_3=0, \quad \beta_1=\frac12; \label{P6rat1}\\
y(x)=\frac{2(\beta_3+\beta_2x)^2-\beta_3-\beta_2x^2}{(2\beta_2+2\beta_3-1)(\beta_3+\beta_2x)}\;,
& \quad & \beta_{\infty}+\beta_1+\beta_2+\beta_3=0, \quad \beta_1=-1.\label{P6rat2}
\end{eqnarray}
As an illustration, we see that the solution obtained in Example \ref{example_1} for $n=1$ is equivalent to $(\ref{P6rat2})$ with $\beta_1=-1$, $\beta_2=\beta_3=1/6$, $\beta_{\infty}=2/3$ by the symmetry $x\mapsto1/x$, $y\mapsto1/y$, $\beta_{\infty}\leftrightarrow\beta_1+1/2$.

It also turns out, as we will see in the next section, that particular Painlev\'e VI equations possess {\it one-parameter families} of rational solutions, not only isolated ones. In our understanding, the emergence of such one-parameter families is not clarified in \cite{Maz11}: on one hand, they occur under the action of particular birational canonical transformations on the degenerate solutions; on the other hand,  solutions (\ref{P6rat1}), (\ref{P6rat2}) are included in one-parameter rational families for particular values of the parameters $\beta_i$'s. This delicate issue is discussed
in Section \ref{OkamotoMazzocco}, which, however, is not of direct relevance to applications of our main study,
where we explain in more detail Okamoto's birational canonical transformations and their action on the degenerate solutions of Painlev\'e VI equations.}
\end{remark}

\subsection{Families of  rational solutions of $P_{\rm VI}$: a sphere with three punctures}
\label{sect_rationalPVI_sphere}

Now we consider the case (c) of Theorem \ref{thm}, that is, the case of $n<0$, $m>0$ continuing to illustrate this theorem with rational solutions of Painlev\'e VI equations. To obtain rational expressions for the $b_i$'s by Theorem \ref{thm_ration} in this case, one requires $1/m\in{\mathbb Z}$, that is, $m=1$. Hence we deal with the Riemann surface $X_x={\mathbb CP}^1$ of the curve
$$
w=z(z-1)(z-x)
$$
punctured at the three points $(0,0), (1,0), (x,0)$. There are two basic cycles on $X_x\setminus\{(0,0),(1,0),(x,0)\}$
and the integration of the triple $w^ndz/z$, $w^ndz/(z-1)$, $w^ndz/(z-x)$ along these very cycles, due to Theorem \ref{thm_conj}, gives us two basic elements $(b^{\rm R}_1(x),b^{\rm R}_2(x),b^{\rm R}_3(x))$ and $(\tilde b^{\rm R}_1(x),\tilde b^{\rm R}_2(x),\tilde b^{\rm R}_3(x))$ in the two-dimensional space of triangular solutions (\ref{sol22}) of the Schlesinger system (\ref{Schl22}), with
$$
\beta_1=\beta_2=\beta_3=\frac n2<0, \quad \beta_{\infty}=-\frac{3n}2.
$$
These basic solutions are rational according to Theorem \ref{thm_ration}, their explicit expressions are presented below. In turn, the pairs $b^{\rm R}_1$,
$\tilde b^{\rm R}_1$ and $b^{\rm R}_2$, $\tilde b^{\rm R}_2$ are basic solutions of the corresponding hypergeometric equations (\ref{hypergeom1}) and (\ref{hypergeom2}), which are thus solvable in rational functions.

Let us take two basic cycles on $X_x\setminus\{(0,0),(1,0),(x,0)\}$ encircling, for example, the points $(a_1,0)=(0,0)$ and $(a_2,0)=(1,0)$ and compute
the corresponding residues of the differentials $w^ndz/z$, $w^ndz/(z-1)$, $w^ndz/(z-x)$. 
 The coordinate representation of these differentials in a local parameter $t_1$ near the point $(a_1,0)$,
according to formula (\ref{omegaram}) with $N=3$, $(a_1,a_2,a_3)=(0,1,x)$ and $m=1$, is of the form
$$
\frac{w^ndz}z=(1-t_1)^{-|n|}\,(x-t_1)^{-|n|}\,\frac{dt_1}{t_1^{|n|+1}}, \qquad \frac{w^ndz}{z-1}=-(1-t_1)^{-|n|-1}\,(x-t_1)^{-|n|}\,\frac{dt_1}{t_1^{|n|}},
$$
$$
\frac{w^ndz}{z-x}=-(1-t_1)^{-|n|}\,(x-t_1)^{-|n|-1}\,\frac{dt_1}{t_1^{|n|}}.
$$
The first differential has therefore the following expansion near $t_1=0$:
$$
\frac1{x^{|n|}}\sum_{j_1=0}^{\infty}{-|n|\choose j_1}(-t_1)^{j_1}\sum_{j_2=0}^{\infty}{-|n|\choose j_2}(-t_1/x)^{j_2}\,\frac{dt_1}{t_1^{|n|+1}},
$$
whence its residue $b_1^{\rm R}(x)$ at $t_1=0$ equals, up to a constant factor of $(-1)^n$,
\begin{equation*}
b_1^{\rm R}(x)=\frac1{x^{2|n|}} \sum_{j=0}^{|n|}{-|n|\choose j} {-|n|\choose |n|-j}x^j\,.
\end{equation*}
 Similarly, for the residues $b_2^{\rm R}(x)$, $b_3^{\rm R}(x)$ of the two remaining differentials, up to the same factor of $(-1)^n$, one has
\begin{eqnarray}
b_2^{\rm R}(x)&=& \frac1{x^{2|n|-1}}\sum_{j=0}^{|n|-1}{-|n|-1\choose j} {-|n|\choose |n|-1-j}x^j\,,\;\;
 \nonumber\\
  \nonumber\\
b_3^{\rm R}(x)&=& \frac1{x^{2|n|}} \sum_{j=0}^{|n|-1} {-|n|\choose j}{-|n|-1\choose |n|-1-j}x^j  \,. \nonumber
\end{eqnarray}
In an analogous way, the local representation of the  three above differentials near the point $(a_2,0)$ in the local parameter $t_2$ has the form
$$
\frac{w^ndz}z=(1+t_2)^{-|n|-1}\,(1-x+t_2)^{-|n|}\,\frac{dt_2}{t_2^{|n|}}, \qquad \frac{w^ndz}{z-1}=(1+t_2)^{-|n|}\,(1-x+t_2)^{-|n|}\,\frac{dt_2}{t_2^{|n|+1}},
$$
$$
\frac{w^ndz}{z-x}=(1+t_2)^{-|n|}\,(1-x+t_2)^{-|n|-1}\,\frac{dt_2}{t_2^{|n|}},
$$
hence their residues at this point are, respectively,
\begin{eqnarray}
\tilde b_1^{\rm R}(x)&=&  \frac{1}{(1-x)^{2|n|-1}} \sum_{j=0}^{|n|-1}{-|n|-1\choose j} {-|n|\choose |n|-1-j}(1-x)^j   \,,
 \nonumber\\
  \nonumber\\
\tilde b_2^{\rm R}(x)&=& \frac{1}{(1-x)^{2|n|}} \sum_{j=0}^{|n|}{-|n|\choose j} {-|n|\choose |n|-j}(1-x)^j    \,,\;\;
 \nonumber\\ \nonumber\\
\tilde b_3^{\rm R}(x)&=& \frac{1}{(1-x)^{2|n|}} \sum_{j=0}^{|n|-1} {-|n|\choose j}{-|n|-1\choose |n|-1-j}(1-x)^j  \,.
\nonumber
\end{eqnarray}

Again, for any $n<0\,,$ according to \eqref{b22} the functions $cb_1^{\rm R}(x)+\tilde b_1^{\rm R}(x)$ and
$c b_3^{\rm R}(x)+\tilde b_3^{\rm R}(x)$, $c\in{\mathbb C}$, give a rational solution to the Painlev\'e VI equation with parameters
\begin{equation*}
(\alpha,\beta,\gamma,\delta)=\left(\frac{(2\beta_\infty-1)^2}2, -2\beta_1^2, \, 2\beta_2^2, \, \frac12-2\beta_3^2\right)=
                             \left(\frac{(3n+1)^2}2, -\frac{n^2}{2}, \, \frac{n^2}{2}, \, \frac{1-n^2}{2}\right),
\end{equation*}
and thus we obtain the following theorem.
\medskip

\begin{theorem}
\label{thm_rat2}
For every negative integer $n$, the functions
\begin{equation*}
y(x)=\frac{x(c\,b_1^{\rm R}(x)+\tilde b_1^{\rm R}(x))}{c\,b_1^{\rm R}(x)+\tilde b_1^{\rm R}(x)+(1-x)(c\,b_3^{\rm R}(x)+\tilde b_3^{\rm R}(x))}, \qquad c\in{\mathbb C},
\end{equation*}
give a one-parameter family of rational solutions of the sixth Painlev\'e equation ${\rm P_{VI}}(\alpha,\beta,\gamma,\delta)$ with the parameters
\begin{equation*}
\alpha=\frac{(3n+1)^2}2, \quad \beta=-\frac{n^2}{2}, \quad \gamma=\frac{n^2}{2}, \quad \delta=\frac{1-n^2}{2}.
\end{equation*}
\end{theorem}

Like in the previous section, the monodromy of the triangular Schlesinger isomonodromic family corresponding to the above
$b_i^{\rm P}$'s is commutative, since the eigenvalues $e^{\pm2\pi{\bf i}\beta_i}=e^{\pm\pi{\bf i}n}$ of each monodromy matrix $M_i$ coincide
(and all $M_i$'s may be chosen triangular).

\begin{example}\label{exratfam}
{\rm Let us compute two basic rational solutions to the hypergeometric equations \eqref{hypergeom1}, \eqref{hypergeom2} with
$\beta_1=\beta_2=\beta_3=n/2<0$ and the corresponding  family of  rational solutions to the  Painlev\'e VI equation in the case $n=-1$, $n=-2$, and $n=-3$.
\begin{enumerate}
\item For $n=-1$, we obtain
\begin{eqnarray*}
b_1^{\rm R}(x)=-\frac{1+x}{x^2}\;, &
\qquad
&\tilde b_1^{\rm R}(x)=\frac{1}{1-x}\;,
\\
b_2^{\rm R}(x)=\frac{1}{x}\;, &
\qquad
&\tilde b_2^{\rm R}(x)=\frac{x-2}{(1-x)^2}\;,
\\
b_3^{\rm R}(x)=\frac{1}{x^2}\;, &
\qquad
&\tilde b_3^{\rm R}(x)=\frac{1}{(1-x)^2}\;,
\end{eqnarray*}
where $b_1^{\rm R}$ and $\tilde b_1^{\rm R}$ satisfy \eqref{hypergeom1} and $b_2^{\rm R}, \;\tilde b_2^{\rm R}$ satisfy \eqref{hypergeom2} with $\beta_1=\beta_2=\beta_3=-1/2$. The corresponding family of rational solutions of the sixth Painlev\'e equation
${\rm P_{VI}}\bigl(2, -\frac{1}{2}, \frac{1}{2}, 0\bigr)\,$ is given by
\begin{equation*}
y(x)= \frac{1}{2} \, \frac{(1-c)x^2+c}{(1-c)x+c}\;, \qquad c\in{\mathbb C}.
\end{equation*}
\item For $n=-2$, we obtain
\begin{eqnarray*}
b_1^{\rm R}(x)=\frac{3+4x+3x^2}{x^4}\;, &
\qquad
&\tilde b_1^{\rm R}(x)= \frac{-5+3x}{(1-x)^3}\;,
\\
b_2^{\rm R}(x)= -\frac{2+3x}{x^3}\;, &
\qquad
&\tilde b_2^{\rm R}(x)=\frac{10-10x+3x^2}{(1-x)^4}\;,
\\
b_3^{\rm R}(x)=-\frac{3+2x}{x^4}\;, &
\qquad
&\tilde b_3^{\rm R}(x)=\frac{-5+2x}{(1-x)^4}\;,
\end{eqnarray*}
where $b_1^{\rm R}$ and $\tilde b_1^{\rm R}$ satisfy \eqref{hypergeom1} and $b_2^{\rm R}, \;\tilde b_2^{\rm R}$ satisfy  \eqref{hypergeom2} with $\beta_1=\beta_2=\beta_3=-1$. The corresponding family of rational solutions of the sixth Painlev\'e equation ${\rm P_{VI}}\bigl(\frac{25}2,-2,2, -\frac32\bigr)\,$ is given by
\begin{equation*}
y(x)= \frac{1}{5} \, \frac{(1-c)x^4(3x-5) + c(3-5x)}{(1-c)x^3(x-2) + c(1-2x)}\;,\qquad c\in{\mathbb C}.
\end{equation*}
\item For $n=-3$, we obtain
\begin{eqnarray*}
b_1^{\rm R}(x)= -\frac{10+18x+18x^2+10x^3}{x^6} \;, &
\qquad
&\tilde b_1^{\rm R}(x)= \frac{28-32x+10x^2}{(1-x)^5}\;,
\\
b_2^{\rm R}(x)= \frac{6+12x+10x^2}{x^5}\;, &
\qquad
&\tilde b_2^{\rm R}(x)=-\frac{-56+84x-48x^2+10x^3}{(1-x)^6}\;,
\\
b_3^{\rm R}(x)= \frac{10+12x+6x^2}{x^6}\;, &
\qquad
&\tilde b_3^{\rm R}(x)= \frac{28-24x+6x^2}{(1-x)^6}\;,
\end{eqnarray*}
where $b_1^{\rm R}$ and $\tilde b_1^{\rm R}$ satisfy \eqref{hypergeom1} and $b_2^{\rm R}, \;\tilde b_2^{\rm R}$ satisfy  \eqref{hypergeom2} with $\beta_1=\beta_2=\beta_3=-3/2$. The corresponding family of
rational solutions of the sixth Painlev\'e equation ${\rm P_{VI}}\bigl(32, -\frac{9}{2}, \frac{9}{2}, -4\bigr)\,$ is given by
\begin{equation*}
y(x)= \frac{1}{4}\frac{(1-c)x^6(14-16x+5x^2)+c(5-16x+14x^2)}{(1-c)x^5(7-7x+2x^2) + c(2-7x+7x^2)}\;, \qquad c\in{\mathbb C}.
\end{equation*}
\end{enumerate}
}
\end{example}

We note that, in the previous section and in the current one, we obtained two essentially different sets of rational solutions of Painlev\'e VI equations as particular cases of our algebro-geometric solutions corresponding to two different Riemann surfaces. Namely, in Section \ref{sect_rationalPVI_torus}, the obtained rational solutions are isolated, whereas the solutions obtained in the current section form a one-parameter family (for a fixed equation).

Concluding these two sections we observe that, outside of the framework of the algebro-geometric approach, one could construct rational solutions of
Painlev\'e VI equations for a much larger set of parameters $\alpha$, $\beta$, $\gamma$, $\delta$ than the above discrete sets of the illustrative Theorems
\ref{thmrat1}, \ref{thm_rat2}. As one may guess, a general hint for this is to search for those values of the parameters $\bf a$, $\bf b$, $\bf c$ for which one or even two basic solutions of the corresponding hypergeometric equation (\ref{hypergeomgen}) are expressed {\it via} truncated hypergeometric power
series and thus reduced to rational functions. We do not detail this general approach here, since it is not related directly to our main study. Just to
compare with formulae (\ref{eq:PailneveVIJacobi1}), (\ref{eq:PailneveVIJacobi2}) of Proposition \ref{prop:rat1},  we formulate the following

\begin{proposition}
For every positive integer $n$ and rather generic values of complex parameters
$\bf b$, $\bf c$, the sixth Painlev\'e equation ${\rm P_{VI}}(\alpha,\beta,\gamma,\delta)$ with the parameters
$$
\alpha=\frac{(n+1)^2}2, \quad \beta=-\frac{(1+{\bf b}-{\bf c})^2}2, \quad \gamma=\frac{(1-n-{\bf c})^2}2, \quad \delta=\frac{1-{\bf b}^2}2,
$$	
possesses a rational solution $y_n(x)=y_n({\bf b},{\bf c},x)$ that is expressed {\rm via} Jacobi polynomials as follows:
\begin{eqnarray*}
{\rm (a)} \qquad y_n(x)&=& \frac{{\bf b}-{\bf c}+1}{{\bf b}+n}+\frac{(n+{\bf c}-1)({\bf c}-{\bf b}-1)}{({\bf b}+n)(n+1)}\,
\frac{P_n^{({\bf c}-2, {\bf b}-{\bf c}-n)}(1-2x)}{P_{n+1}^{({\bf c}-2, {\bf b}-{\bf c}-n)}(1-2x)}; \\
{\rm (b)} \qquad y_n(x)&=& \frac{{\bf b}-{\bf c}+1}{{\bf b}+n}-\frac{(n+{\bf c}-1)({\bf c}-{\bf b}-1)}{({\bf b}-2)({\bf b}+n)(n+1)}\,
\frac{\frac d{dx}P_{n+1}^{({\bf c}-3, {\bf b}-{\bf c}-n-1)}(1-2x)}{P_{n+1}^{({\bf c}-2, {\bf b}-{\bf c}-n)}(1-2x)}.
\end{eqnarray*}
\end{proposition}

 \subsection{Around Okamoto's birational transformations and classification of $P_{VI}$ rational solutions}
\label{OkamotoMazzocco}

Birational canonical transformations (of the {\it first kind}) of Painlev\'e VI equations, as they were defined by K.\,Okamoto \cite{Ok2}, act on the pair,
the initial unknown $y$ and its conjugated momentum $p$, with respect to which the sixth Painlev\'e equation ${\rm P_{VI}}(\alpha,\beta,\gamma,\delta)$
can be rewritten as a first order system:
\begin{eqnarray}\label{Psyst}
\left\{\begin{array}{ll}y'=&\frac{y(y-1)(y-x)}{x(x-1)}\left(2p-\frac{2\beta_3-1}{y-x}-\frac{2\beta_1}y-\frac{2\beta_2}{y-1}\right) \\
\\
p'=&-\frac1{x(x-1)}\Bigl([3y^2-2(x+1)y+x]p^2+[(2-4\beta_1-4\beta_2-4\beta_3)y+\\
\\ &+2\beta_1+2\beta_3-1+(2\beta_1+2\beta_2)x]p+\varkappa\Bigr),
\end{array}\right.
\end{eqnarray}
where $\varkappa=(\beta_1+\beta_2+\beta_3-\beta_{\infty})(\beta_1+\beta_2+\beta_3+\beta_{\infty}-1)$.

Introducing new parameters
$$
{\rm b_1}=\beta_1+\beta_2, \qquad {\rm b_2}=\beta_1-\beta_2, \qquad {\rm b_3}=\beta_3+\beta_{\infty}-1, \qquad {\rm b_4}=\beta_3-\beta_{\infty},
$$
Okamoto defines the following affine transformations on their space ${\mathbb C}^4$:
\begin{eqnarray*}
w_1: && ({\rm b_1},{\rm b_2},{\rm b_3},{\rm b_4})\mapsto({\rm b_2},{\rm b_1},{\rm b_3},{\rm b_4}), \\
w_2: && ({\rm b_1},{\rm b_2},{\rm b_3},{\rm b_4})\mapsto({\rm b_1},{\rm b_3},{\rm b_2},{\rm b_4}), \\
w_3: && ({\rm b_1},{\rm b_2},{\rm b_3},{\rm b_4})\mapsto({\rm b_1},{\rm b_2},{\rm b_4},{\rm b_3}), \\
w_4: && ({\rm b_1},{\rm b_2},{\rm b_3},{\rm b_4})\mapsto(-{\rm b_2},-{\rm b_1},{\rm b_3},{\rm b_4}),
\end{eqnarray*}
and $w_0: ({\rm b_1},{\rm b_2},{\rm b_3},{\rm b_4})\mapsto({\rm b_1},{\rm b_2},-{\rm b_4}-1,-{\rm b_3}-1)$. It turns out that each transformation $w$
is induced by a birational transformation $(y,p)\mapsto(y_w,p_w)$ of the Painlev\'e system (\ref{Psyst}) {\it via} the formula
\begin{equation}\label{biratform}
{\bf F}[{\rm b}]\left(\begin{array}{c} y \\ y(y-1)p \end{array}\right)+{\bf g}[{\rm b}]=
{\bf F}[w({\rm b})]\left(\begin{array}{c} y_w \\ y_w(y_w-1)p_w \end{array}\right)+{\bf g}[w({\rm b})],
\end{equation}
where ${\rm b}=({\rm b_1},{\rm b_2},{\rm b_3},{\rm b_4})$ and
\begin{equation}\label{Fg}
{\bf F}[{\rm b}]=\left(\begin{array}{ll} -h+\sigma'_2[{\rm b}] & -{\rm b_3}-{\rm b_4} \\
\sigma'_1[{\rm b}]h-\sigma'_3[{\rm b}] & -h+{\rm b_3}{\rm b_4} \end{array}\right), \qquad
{\bf g}[{\rm b}]=\left(\begin{array}{c} -\frac12\sigma_2[{\rm b}] \\ -\frac12\sigma_1[{\rm b}]h+\frac12\sigma_3[{\rm b}] \end{array}\right).
\end{equation}
In the above formulae, $\sigma_k[{\rm b}]$ denotes the elementary symmetric polynomial of degree $k$ in four variables $\rm b_1$, $\rm b_2$, $\rm b_3$,
$\rm b_4$, and $\sigma'_k[{\rm b}]$ denotes the elementary symmetric polynomial of degree $k$ in three variables $\rm b_1$, $\rm b_3$, $\rm b_4$. The
polynomial $h=h(y,p)$ is given by the formula
$$
h=-y(y-1)p^2+\bigl(2{\rm b_1}y-({\rm b_1}+{\rm b_2})\bigr)p-{\rm b}_1^2.
$$
\begin{remark}
{\rm Originally, the transformation $w_4$ was defined by Okamoto in the form
$$
w_4: ({\rm b_1},{\rm b_2},{\rm b_3},{\rm b_4})\mapsto({\rm b_1},{\rm b_2},-{\rm b_4},-{\rm b_3}).
$$
Note that there is a misprint in the image of this $w_4$, where the last two coordinates are $-{\rm b_3}$, $-{\rm b_4}$ in \cite{Ok2}. Another misprint
in \cite{Ok2} is the absence of the factor $1/2$ in the second coordinate of the vector ${\bf g}[{\rm b}]$ in (\ref{Fg}).
}
\end{remark}

As can be easily seen, the birational transformation associated with $w_3$ does not change $y$ nor $p$, since
${\bf F}[w({\rm b})]={\bf F}[{\rm b}]$ and ${\bf g}[w({\rm b})]={\bf g}[{\rm b}]$ in this case. The birational transformations associated with $w_1$,
$w_4$, and $w_0$ do not change $y$ but change $p$ (they correspond to the change of sign $\beta_2\leftrightarrow-\beta_2$,
$\beta_1\leftrightarrow-\beta_1$, and $\beta_3\leftrightarrow-\beta_3$, respectively).

Birational transformations associated with $w_2$ or with those containing $w_2$ as a factor, change  both $y$ and $p$, and they are of a particular interest
for us. We will study the action of $w_1w_2w_1$ on the {\it degenerate} solutions of the sixth Painlev\'e equation ${\rm P_{VI}}(\alpha,\beta,\gamma,\delta)$. The latter are:
\begin{itemize}
\item[i)] $y(x)\equiv\infty$ for $\alpha=0$ (that is, for $\beta_{\infty}=1/2$);
\item[ii)] $y(x)\equiv0$ for $\beta=0$ (that is, for $\beta_1=0$);
\item[iii)] $y(x)\equiv1$ for $\gamma=0$ (that is, for $\beta_2=0$);
\item[iv)] $y(x)\equiv x$ for $\delta=1/2$ (that is, for $\beta_3=0$).
\end{itemize}
This set is invariant under the action of the symmetries (birational transformations of the {\it second kind}, as they
change the independent variable $x$)
$$
{\rm a)}\; x\mapsto 1-x, \quad y\mapsto 1-y, \quad \beta_1\leftrightarrow\beta_2; \qquad {\rm b)}\; x\mapsto\frac1x, \quad
y\mapsto\frac1y, \quad \beta_{\infty}\leftrightarrow\beta_1+\frac12;
$$
$$
{\rm c)}\; x\mapsto \frac x{x-1}, \quad y\mapsto\frac{y-x}{1-x}, \quad \beta_1\leftrightarrow\beta_3\,.
$$
Note that for any solution $y(x)$ different from i)--iv), its conjugated momentum $p(x)$ is uniquely determined by
the first equation of the Painlev\'e system (\ref{Psyst}), in particular, $p$ is rational if $y$ is. On the other hand, for each of the solutions i)--iv),
its conjugated momentum is a {\it one-parameter family} of solutions of the corresponding Riccati equation coming from the second equation of (\ref{Psyst}). Therefore for such a pair $(y,p)$, the image $y_w$ of $y$ under the birational transformation associated with
$$
w=w_1w_2w_1: ({\rm b_1},{\rm b_2},{\rm b_3},{\rm b_4})\mapsto({\rm b_3},{\rm b_2},{\rm b_1},{\rm b_4}),
$$
can also be a {\it one-parameter family} of solutions of the corresponding sixth Painlev\'e equation. Let us explain this in more detail in the case of
the sixth Painlev\'e equation ${\rm P_{VI}}(\alpha,0,\gamma,\delta)$ possessing the degenerate solution $y\equiv0$.

For $\beta_1=0$, one has ${\rm b_1}=-{\rm b_2}$ and the polynomial $h$ is equal to
$$
h=-y(y-1)p^2+2{\rm b_1}yp-{\rm b}_1^2.
$$
Taking into consideration the equalities ${\bf g}[w({\rm b})]={\bf g}[{\rm b}]$ and $\sigma'_k[w({\rm b})]=\sigma'_k[{\rm b}]$
for $w=w_1w_2w_1$, one obtains from (\ref{biratform}) the formula
\begin{eqnarray*}
\left(\begin{array}{c} y_w \\ y_w(y_w-1)p_w \end{array}\right)=
\frac1{h+{\rm b}_1^2}\left(\begin{array}{ll} h+{\rm b}_1^2 & {\rm b_3}-{\rm b_1} \\
0 & h+{\rm b}_3^2 \end{array}\right)\left(\begin{array}{c} y \\ y(y-1)p \end{array}\right)
\end{eqnarray*}
(see Example 2.1 on p. 356 in \cite{Ok2}), which implies
$$
y_w=y+\frac{({\rm b}_3-{\rm b}_1)(y-1)}{-(y-1)p+2{\rm b}_1}, \qquad
p_w=\frac{(y-1)(-y(y-1)p^2+2{\rm b}_1yp-{\rm b}_1^2+{\rm b}_3^2)}{-(y-1)p+2{\rm b}_1}\frac1{y_w(y_w-1)}.
$$

The above formulae give explicitly the action of the birational canonical transformation associated with $w=w_1w_2w_1$ on the Painlev\'e system
(\ref{Psyst}) with $\beta_1={\rm b}_1+{\rm b}_2=0$. Note that $\det{\bf F}[w(\rm b)]=0$ for $y=0$ since in this case $h+{\rm b}_1^2=0$, but the final
expressions for $y_w$, $p_w$ are defined also for $y=0$. We thus have a prolongation
of the birational canonical transformation to the degenerate solution $y\equiv0$ of ${\rm P_{VI}}(\alpha,0,\gamma,\delta)$:
\begin{equation}\label{birat0}
(0,p)\mapsto\Bigl(\frac{{\rm b}_1-{\rm b}_3}{p+2{\rm b}_1},-\frac{({\rm b}_1+{\rm b}_3)(p+2{\rm b}_1)}{p+{\rm b}_1+{\rm b}_3}\Bigr),
\end{equation}
where $p$ is the general solution of the Riccati equation
\begin{equation}\label{Riccati}
-x(x-1)p'=xp^2+(2{\rm b}_1x+{\rm b}_3+{\rm b}_4)p+({\rm b}_1+{\rm b}_3)({\rm b}_1+{\rm b}_4).
\end{equation}
Therefore, if the parameters in equation \eqref{Riccati} were such that its general solution was a rational function, we would obtain the transformation
of the {\it degenerate} solution $y\equiv0$ of ${\rm P_{VI}}(\alpha,0,\gamma,\delta)$ to a {\it one-parameter family} of rational solutions of the
transformed Painlev\'e VI equation, under the action provided by (\ref{birat0}). We give the following example.

\begin{example}\label{exOk}
{\rm

Consider the set of parameters ${\rm b}=({\rm b_1},{\rm b_2},{\rm b_3},{\rm b_4})=(-1,1,0,1)$ and, consequently,
$$
\beta_1=\frac{{\rm b_1}+{\rm b_2}}2=0, \quad \beta_2=\frac{{\rm b_1}-{\rm b_2}}2=-1, \quad \beta_3=\frac{{\rm b_3}+{\rm b_4}+1}2=1,
\quad \beta_{\infty}=\frac{{\rm b_3}-{\rm b_4}+1}2=0.
$$	
The corresponding sixth Painlev\'e equation ${\rm P_{VI}}\bigl(\frac12,0,2,-\frac32\bigr)$ possesses the degenerate solution $y\equiv0$, whose
conjugated momentum $p$ is the general solution of the Riccati equation (\ref{Riccati})
$$
-x(x-1)p'=xp^2+(-2x+1)p,
$$
that is,
$$
p(x)=\frac{2x(x-1)}{x^2+c}, \qquad c\in{\mathbb C}.
$$
Since $w_1w_2w_1({\rm b})=(0,1,-1,1)$, under the action of the associated birational transformation the solution $y\equiv0$ of
${\rm P_{VI}}\bigl(\frac12,0,2,-\frac32\bigr)$ is mapped, according to (\ref{birat0}), to the one-parameter family of rational solutions
\begin{equation}\label{ratfam1}
y_w(x)=\frac{-1}{\frac{2x(x-1)}{x^2+c}-2}=\frac12\,\frac{x^2+c}{x+c}
\end{equation}
of the corresponding sixth Painlev\'e equation ${\rm P_{VI}}\bigl(2,-\frac12,\frac12,0\bigr)$. This family has been already obtained in Example
\ref{exratfam}.
}
\end{example}

Concluding this section we note that it would also be natural to call such one-parameter families of rational solutions of Painlev\'e VI equations
{\it degenerate}, as they are obtained from the degenerate solutions. They do not participate in Mazzocco's classification of rational solutions.
On the other hand, Mazzocco's basic rational solutions (\ref{P6rat1}), (\ref{P6rat2}) themselves, for some values of the parameters $\beta_i$'s,
can belong to one-parameter families of rational solutions of the corresponding Painlev\'e VI equations. For example, when $\beta_1=1/2$, $\beta_2=-1/2$,
$\beta_3=1/2$, $\beta_{\infty}=-1/2$, solution (\ref{P6rat1}) of ${\rm P_{VI}}\bigl(2,-\frac12,\frac12,0\bigr)$ is $y(x)=x/2$, which belongs
to the family (\ref{ratfam1}) and thus can be obtained from the degenerate solution $y\equiv0$  of ${\rm P_{VI}}\bigl(\frac12,0,2,-\frac32\bigr)$ {\it via} a birational transformation. Similarly, when $\beta_1=\beta_2=\beta_3=-1$, $\beta_{\infty}=3$,  solution (\ref{P6rat2}) of
${\rm P_{VI}}\bigl(\frac{25}2,-2,2,-\frac32\bigr)$ is $y(x)=\frac15\frac{3x^2+4x+3}{1+x}$, which belongs to the second rational family of Example \ref{exratfam} (formally, it corresponds to the value $c=\infty$ of the family parameter).

The reasoning above raises the following questions: (i) For which values of the parameters $\beta_i$'s the corresponding basic rational solution (\ref{P6rat1}) or
(\ref{P6rat2}) is isolated and for which values it belongs to a one-parameter rational family, thus being the candidate for being birationally
equivalent to a degenerate solution? (ii) Are there other one-parameter rational families beside degenerate ones? This shows, in our understanding, that the problem of the classification of rational solutions
is not completely closed.

\begin{remark}
{\rm We also mention the paper \cite{YL} with the classification of rational solutions of Painlev\'e VI equations. However, it was observed in \cite{Alm1}
that Theorem 4.2. from \cite{YL} states that $y(x)$ is a non-constant rational solution of the sixth Painlev\'e equation
${\rm P_{VI}}(\alpha,\beta,\gamma,\delta)$ if and only if its conjugated momentum $p\equiv0$, that is, if and only if $y(x)$ solves the corresponding
Riccati equation (the first equation of the Painlev\'e system (\ref{Psyst}) with $p\equiv0$).
This is not always the case, as we could see in Example \ref{exOk}: the rational solution $y_w(x)$ given by (\ref{ratfam1}) of the sixth Painlev\'e
equation ${\rm P_{VI}}\bigl(2,-\frac12,\frac12,0\bigr)$ solves the algebraic first order ODE of the third degree in $y$ rather than the Riccati equation,
since the conjugated momentum $p_w(x)$ of $y_w(x)$, according to (\ref{birat0}), is
$$
p_w(x)=\frac{p(x)-2}{p(x)-1}\not\equiv0, \qquad p(x)=\frac{2x(x-1)}{x^2+c}.
$$
}
\end{remark}

\section{Application to  Garnier systems}

Here we consider Garnier systems ${\cal G}_M(\theta)$ (a multidimensional generalization of Painlev\'e VI equations) depending on
$M+3$ complex parameters $\theta_1,\ldots,\theta_{M+2},\theta_{\infty}$. These are completely integrable PDEs systems of second order
\cite{Gar1}, \cite{Gar2}. They can be written in a Hamiltonian form obtained by K.\,Okamoto \cite{Ok},
\begin{equation}\label{garnier}
\frac{\partial u_i}{\partial a_j}=\frac{\partial H_j}{\partial v_i}, \qquad
\frac{\partial v_i}{\partial a_j}=-\frac{\partial H_j}{\partial u_i}, \qquad i,j=1,\ldots,M,
\end{equation}
for  the unknown functions $(u,v)=(u_1,\ldots,u_M,v_1,\ldots,v_M)$ of the variable $a=(a_1,\ldots,a_M)$, where the Hamiltonians
$H_j=H_j(a,u,v,\theta)$ are rational functions of their arguments (see also \cite{IKSY} and Example \ref{ex_garnier1}  below).

Let us recall how the Garnier system is determined by the Schlesinger system for $M+2$ {\it traceless} $(2\times2)$-matrices
$B^{(1)}(a),\ldots,B^{(M+2)}(a)$ depending on  the variable $a$ (here $a_{M+1}=0$, $a_{M+2}=1$ are fixed) which belongs to a disc $D$ of the space
$({\mathbb C}\setminus\{0,1\})^M\setminus\bigcup_{i\ne j}\{a_i=a_j\}$.

Let $\pm\beta_i$ be the eigenvalues of the matrix $B^{(i)}(a)=\bigl(b_i^{kl}(a)\bigr)_{1\leqslant k,l\leqslant2}$, $i=1,\ldots,M+2$, and
$$
\sum_{i=1}^{M+2}B^{(i)}(a)={\rm diag}(-\beta_{\infty},\beta_{\infty}).
$$
Since $\sum_{i=1}^{M+2}b_i^{12}(a)\equiv0$, the numerator of the fraction
$$
\sum_{i=1}^{M+2}\frac{b_i^{12}(a)}{z-a_i}
$$
is a polynomial of degree $M$ in $z$. If one denotes  its zeros by $u_1(a),\ldots,u_M(a)$ and defines
\begin{equation}\label{v}
v_j(a)=\sum_{i=1}^{M+2}\frac{b_i^{11}(a)+\beta_i}{u_j(a)-a_i}, \qquad j=1,\ldots,M,
\end{equation}
then the pair $(u,v)=(u_1,\ldots,u_M,v_1,\ldots,v_M)$ satisfies the Garnier system (\ref{garnier}) with parameters
$$
(\theta_1,\ldots,\theta_{M+2},\theta_{\infty})=(2\beta_1,\ldots,2\beta_{M+2},2\beta_{\infty}-1)
$$
(see proof of Prop. 3.1 in \cite{Ok} or \cite[Cor. 6.2.2]{IKSY}).  Since the functions $u_1(a),\ldots,u_M(a)$ depend on the $b_i^{12}$'s algebraically, for $M>1$ they are, in general, not meromorphic on the universal cover $Z$ of the space $({\mathbb C}\setminus\{0,1\})^M\setminus\bigcup_{i\ne j}\{a_i=a_j\}$. However, some information concerning the elementary symmetric polynomials in the coordinates  $u_1,\ldots,u_M$ can be obtained in this context (see, for example \cite{GV}).

As we have seen in Section \ref{Painleve}, solutions of the Schlesinger system for {\it triangular} traceless $(2\times2)$-matrices depending on
$ M=1$ variable always lead to solutions of the corresponding sixth Painlev\'e equation that are expressed rationally {\it via} a
logarithmic derivative of solutions of a hypergeometric linear ODE. This fact admits a generalization for the
multivariable case of {\it triangular} traceless $(2\times2)$-matrices depending on $M>1$ variables that solve the Schlesinger system:
they lead to solutions of the corresponding Garnier system that are expressed algebraically {\it via} logarithmic derivatives
of solutions of a Lauricella hypergeometric PDE. Before exposing this in more detail, let us recall that the latter is
a  system of linear PDEs of the second order of the form
\begin{eqnarray*}
(1-a_i)\sum_{j=1}^Ma_j\frac{\partial^2u}{\partial a_i\partial a_j}+(\varkappa-(\alpha+1)a_i)\frac{\partial u}{\partial a_i}-
\mu_i\sum_{j=1}^Ma_j\frac{\partial u}{\partial a_j}-\alpha\mu_iu=0, \qquad i=1,\ldots,M,\\
(a_i-a_j)\frac{\partial^2u}{\partial a_i\partial a_j}+\mu_i\frac{\partial u}{\partial a_j}-\mu_j\frac{\partial u}{\partial a_i}=0,
\qquad i,j=1,\ldots,M,
\end{eqnarray*}
for  the unknown function $u$ of $M$ variables $a_1,\ldots,a_M$, where $\alpha,\mu_1,\ldots,\mu_M,\varkappa$ are complex parameters.
Its solution space is $(M+1)$-dimensional, as follows from the proof of Prop. 9.1.4 in \cite{IKSY}. Now, in
 the  triangular case,  a solution
\begin{equation}\label{res_garnier}
B^{(i)}(a) = \left(\begin{array}{cc}\beta_i & b_i(a) \\0 & -\beta_i\end{array}\right)\,, \qquad i=1,\ldots,M+2, \qquad
\sum_{i=1}^{M+2}B^{(i)}(a)={\rm diag}(-\beta_{\infty},\beta_{\infty}),
\end{equation}
of the Schlesinger system determines the polynomial
$$
P_M(z,a)=(z-a_1)\ldots(z-a_{M+2})\sum_{i=1}^{M+2}\frac{b_i(a)}{z-a_i}
$$
of degree $M$ in $z$ with zeros $u_1(a),\ldots,u_M(a)$. Then due to (\ref{v}), the pair
\begin{equation}
\label{hyp_sol_garnier}
(u,v)^{\varepsilon} = (u_1,\ldots,u_M,v_1^{\varepsilon},\ldots,v_M^{\varepsilon}),
\end{equation}
where
\begin{equation}
v_j^{\varepsilon}(a)=\sum_{i=1}^{M+2}\frac{(1+\varepsilon_i)\beta_i}{u_j(a)-a_i}, \qquad\mbox{with}\qquad
 \varepsilon=(\varepsilon_1,\ldots,\varepsilon_{M+2})\in\{\pm1\}^{M+2}, \nonumber
\end{equation}
satisfies the Garnier system (\ref{garnier}) with parameters
\begin{equation}\label{parameters}
(\theta_1,\ldots,\theta_{M+2},\theta_{\infty})=(2\varepsilon_1\beta_1,\ldots,2\varepsilon_{M+2}\beta_{M+2},2\beta_{\infty}-1).
\end{equation}

Then,  introducing new independent variables $t=(t_1,\dots,t_M)$ with
$$
t_i=\frac{a_i}{a_i-1}, \qquad i=1,\ldots,M,
$$
and functions
$$
q_i(t)=a_i\frac{(a_i-u_1)\ldots(a_i-u_M)}{\prod_{j=1,j\ne i}^{M+2}(a_i-a_j)}, \qquad i=1,\ldots,M,
$$
one has the following expressions for the latter:
$$
q_i(t)=\frac{t_i(t_i-1)}{2\beta_{\infty}-1}\left(\frac{-2\beta_i}{t_i-1}+\frac1f\frac{\partial f}{\partial t_i}\right),
$$
where $f$ is a solution of the Lauricella hypergeometric equation with parameters
$$
(\alpha,\mu_1,\ldots,\mu_M,\varkappa)=\bigl(1+2\beta_{M+2},-2\beta_1,\ldots,-2\beta_M,-2\sum_{j=1}^{M+1}\beta_j\bigl)
$$
(see \cite[Th. 9.2.1]{IKSY}).

After mentioning these general relations between triangular Schlesinger $(2\times2)$-systems and Lauricella hypergeometric equations,
we pass to  the particular case we consider,  when  the eigenvalues in (\ref{res_garnier}) are given by
$$
\beta_1=\ldots=\beta_{M+2}=\frac n{2m}, \qquad \beta_{\infty}=-\frac{(M+2)n}{2m},
$$
with any coprime integers $n>0$, $m>1$ or $n<0$, $m>0$.  Applying Theorems \ref{thm}  and \ref{thm_conj}  we obtain algebro-geometric expressions
for an $(M+1)$-parameter family of solutions of the corresponding triangular Schlesinger $(2\times2)$-system
(and thus, for an $(M+1)$-parameter family of the coefficients of the polynomial $P_M(z,a)$):
\begin{equation}\label{alggeom_garnier}
b_i(a)= c_1\oint_{\gamma_1} \frac{w^{n} dz}{z-a_i}+\ldots+c_{M+1}\oint_{\gamma_{M+1}} \frac{w^{n} dz}{z-a_i}\,,  \qquad i=1,\ldots,M+2,
\end{equation}
where  $\gamma_1,\ldots,\gamma_{M+1}$ are suitable  closed contours on the Riemann surface $X_a$ of the curve
$$
w^m=z(z-1)(z-a_1)\ldots(z-a_M)
$$
(or, on the $X_a$  with punctures at the poles of the differentials $w^ndz/z$, $w^ndz/(z-1)$, $w^ndz/(z-a_1), \ldots, w^ndz/(z-a_M)$,  depending on which of the cases (a), (b), (c) of Theorem \ref{thm} holds).
These expressions lead to an $M$-parameter  families of algebro-geometric solutions \eqref{hyp_sol_garnier} of
the Garnier systems with parameters \eqref{parameters}:
\begin{equation*}
(\theta_1,\ldots,\theta_{M+2},\theta_{\infty})=\Bigl(\pm\frac nm,\ldots,\pm\frac nm,-\frac{(M+2)n}m-1\Bigr)\,,
\end{equation*}
the signs being independent.

Like for the Painlev\'e VI equations, let us study in more detail the cases when Theorems \ref{thm_2} and \ref{thm_ration} can
be applied to obtain polynomial and rational expressions for $b_i$'s and, as a consequence, algebraic solutions of particular Garnier systems.

\subsection{Algebraic solutions of Garnier systems:  a surface of positive genus with $m$ punctures}

 In this section we consider the case of $n>0$, $m>1$. The requirement $s/m\in{\mathbb Z}$ of Theorem \ref{thm_2} for $s=(m,M+2)$ implies that $s=m$ and $m$ is a divisor of the integer $M+2$. Hence we deal with the Riemann surface $X_a$ of the curve
$$
w^m=z(z-1)(z-a_1)\ldots(z-a_M)
$$
punctured at $m$ points $P_1, \ldots, P_m$ at infinity. The genus of $X_a$ equals
$$
g=\frac12\bigl((m-1)(M+1)-m+1\bigr)=\frac12(m-1)M,
$$
thus there are $(m-1)(M+1)$ basic cycles on $X_a\setminus\{P_1,\ldots,P_m\}$.

Further,  formula (\ref{polynsol}), where $d=n$, $s=m$, $M_1=(M+2)/m$ and the role of
$N$ being played by $M+2$, gives us the following polynomial solutions (\ref{res_garnier}) of the Schlesinger $(2\times2)$-system in the case $\beta_1=\ldots=\beta_{M+2}=n/2m>0$:
\begin{equation}\label{Schl_polyn}
b_i(a)=\sum_{k_1+\ldots+k_M+k_{M+2}+q=M_1n}(-1)^q{n/m\choose k_1}\ldots{n/m\choose k_M}{n/m\choose k_{M+2}}a_1^{k_1}\ldots a_M^{k_M}a_i^q
\end{equation}
(recall that $a_{M+1}=0$, $a_{M+2}=1$). Hence, the coefficients of the corresponding polynomial $P_M(z,a)$ are also polynomials (in $a_1,\ldots,a_M$)
in this case, and thus we come to the following assertion concerning algebraic solutions of Garnier systems.
\medskip

\begin{theorem}\label{thm_5}
For any coprime integers $n>0$, $m>1$ such that $m$ is a divisor of the integer $M+2$, the Garnier system ${\cal G}_M (\theta)$ with parameters
$$
(\theta_1,\ldots,\theta_{M+2},\theta_{\infty})=\Bigl(\pm\frac nm,\ldots,\pm\frac nm,-\frac{(M+2)n}m-1\Bigr)
$$
$($the signs are independent$)$ possesses an algebraic solution, which can be computed explicitly.
\end{theorem}
\medskip

\begin{example} \label{ex_garnier1}
{\rm Consider some examples of bivariate Garnier systems in the variables $a_1$, $a_2$. The system
${\cal G}_2(\theta_1,\theta_2,\theta_3,\theta_4,\theta_{\infty})$ has the form
\begin{eqnarray*}
\frac{\partial u_1}{\partial a_1}=\frac{\partial H_1}{\partial v_1}, \qquad \frac{\partial u_1}{\partial a_2}=\frac{\partial H_2}{\partial v_1}, & \quad &
\frac{\partial u_2}{\partial a_1}=\frac{\partial H_1}{\partial v_2}, \qquad \frac{\partial u_2}{\partial a_2}=\frac{\partial H_2}{\partial v_2},\\
\frac{\partial v_1}{\partial a_1}=-\frac{\partial H_1}{\partial u_1}, \qquad \frac{\partial v_1}{\partial a_2}=-\frac{\partial H_2}{\partial u_1}, & \quad &
\frac{\partial v_2}{\partial a_1}=-\frac{\partial H_1}{\partial u_2}, \qquad \frac{\partial v_2}{\partial a_2}=-\frac{\partial H_2}{\partial u_2},
\end{eqnarray*}
with the Hamiltonians
\begin{eqnarray*}
H_1=-\frac{\Lambda(a_1)}{T'(a_1)}\sum_{j=1}^2\frac{T(u_j)}{(u_j-a_1)\Lambda'(u_j)}\left[v_j^2-\Bigl(\frac{\theta_1-1}{u_j-a_1}+
\frac{\theta_2}{u_j-a_2}+\frac{\theta_3}{u_j}+\frac{\theta_4}{u_j-1}\Bigr)v_j+\frac{\varkappa}{u_j(u_j-1)}\right], \\
H_2=-\frac{\Lambda(a_2)}{T'(a_2)}\sum_{j=1}^2\frac{T(u_j)}{(u_j-a_2)\Lambda'(u_j)}\left[v_j^2-\Bigl(\frac{\theta_1}{u_j-a_1}+
\frac{\theta_2-1}{u_j-a_2}+\frac{\theta_3}{u_j}+\frac{\theta_4}{u_j-1}\Bigr)v_j+\frac{\varkappa}{u_j(u_j-1)}\right],
\end{eqnarray*}
where $\varkappa=\frac14\bigl((\theta_1+\theta_2+\theta_3+\theta_4-1)^2-\theta_{\infty}^2\bigr)$,
$$
\Lambda(x)=(x-u_1)(x-u_2), \qquad T(x)=x(x-1)(x-a_1)(x-a_2).
$$
The polynomial $P_M(z,a)=P_2(z,a_1,a_2)$ equals
$$
P_2(z,a_1,a_2)=\bigl(b_4+a_1b_1+a_2b_2\bigr)z^2+\bigl(a_1a_2(b_3+b_4)+a_1(b_2+b_3)+a_2(b_1+b_3)\bigr)z-a_1a_2b_3
$$
in this case. As $M+2=4$, there are two divisors of $M+2$: $m=2$ and $m=4$.

\begin{enumerate}
\item Let $m=2$ and $n=1$. Then $M_1=(M+2)/m=2$ and, due to (\ref{Schl_polyn}),
\begin{eqnarray*}
b_1(a_1,a_2)&=&\sum_{k_1+k_2+k_4+q=2}(-1)^q{1/2\choose k_1}{1/2\choose k_2}{1/2\choose k_4}a_1^{k_1+q}a_2^{k_2}=\\
&=&3a_1^2-2a_1a_2-a_2^2-2a_1+2a_2-1, \\
b_2(a_1,a_2)&=&\sum_{k_1+k_2+k_4+q=2}(-1)^q{1/2\choose k_1}{1/2\choose k_2}{1/2\choose k_4}a_1^{k_1}a_2^{k_2+q}=\\
&=&3a_2^2-2a_1a_2-a_1^2+2a_1-2a_2-1, \\
b_3(a_1,a_2)&=&\sum_{k_1+k_2+k_4=2}{1/2\choose k_1}{1/2\choose k_2}{1/2\choose k_4}a_1^{k_1}a_2^{k_2}=\\
&=&-a_1^2+2a_1a_2-a_2^2+2a_1+2a_2-1, \\
b_4(a_1,a_2)&=&\sum_{k_1+k_2+k_4+q=2}(-1)^q{1/2\choose k_1}{1/2\choose k_2}{1/2\choose k_4}a_1^{k_1}a_2^{k_2}=\\
&=&-a_1^2+2a_1a_2-a_2^2-2a_1-2a_2+3
\end{eqnarray*}
(up to a common constant factor $1/8$). The corresponding polynomial $P_2(z,a_1,a_2)$ defines an algebraic function,
two branches $u_1(a_1,a_2)$, $u_2(a_1,a_2)$ of which give us the algebraic solution
$$
(u_1,u_2,v_1^{\varepsilon},v_2^{\varepsilon}), \quad v_j^{\varepsilon}(a_1,a_2)=
\frac14\Bigl(\frac{1+\varepsilon_1}{u_j-a_1}+\frac{1+\varepsilon_2}{u_j-a_2}+\frac{1+\varepsilon_3}{u_j}+\frac{1+\varepsilon_4}{u_j-1}\Bigr),
\quad \varepsilon_i\in\{\pm1\},
$$
of the Garnier system ${\cal G}_2\bigl(\frac{\varepsilon_1}2,\frac{\varepsilon_2}2,\frac{\varepsilon_3}2,\frac{\varepsilon_4}2,-3\bigr)$.

\item Let $m=4$ and $n=1$. Then $M_1=(M+2)/m=1$ and, due to (\ref{Schl_polyn}),
\begin{eqnarray*}
b_1(a_1,a_2)&=&\sum_{k_1+k_2+k_4+q=1}(-1)^q{1/4\choose k_1}{1/4\choose k_2}{1/4\choose k_4}a_1^{k_1+q}a_2^{k_2}=
-3a_1+a_2+1, \\
b_2(a_1,a_2)&=&\sum_{k_1+k_2+k_4+q=1}(-1)^q{1/4\choose k_1}{1/4\choose k_2}{1/4\choose k_4}a_1^{k_1}a_2^{k_2+q}=
a_1-3a_2+1, \\
b_3(a_1,a_2)&=&\sum_{k_1+k_2+k_4=1}{1/4\choose k_1}{1/4\choose k_2}{1/4\choose k_4}a_1^{k_1}a_2^{k_2}=
a_1+a_2+1, \\
b_4(a_1,a_2)&=&\sum_{k_1+k_2+k_4+q=1}(-1)^q{1/4\choose k_1}{1/4\choose k_2}{1/4\choose k_4}a_1^{k_1}a_2^{k_2}=
a_1+a_2-3
\end{eqnarray*}
(up to a common constant factor $-1/4$). Now the corresponding polynomial $P_2(z,a_1,a_2)$ similarly determines the algebraic solutions of the Garnier systems ${\cal G}_2\bigl(\pm\frac14,\pm\frac14,\pm\frac14,\pm\frac14,-2\bigr)$.
\end{enumerate}
}
\end{example}

\subsection{Families of algebraic solutions of Garnier systems: a sphere with $M+2$ punctures}

In this section we consider the case of $n<0$, $m>0$ continuing to study algebraic solutions of Garnier systems. The requirement $1/m\in{\mathbb Z}$ of Theorem \ref{thm_ration} implies that $m=1$ and we deal with the Riemann surface $X_a={\mathbb CP}^1$ of the curve
$$
w=z(z-1)(z-a_1)\ldots(z-a_M)
$$
punctured at the points $(a_1,0),\ldots,(a_M,0),(0,0),(1,0)$.

There are $M+1$ basic cycles on $X_a\setminus\{(a_1,0),\ldots,(a_M,0),(0,0),(1,0)\}$
and the integration of the vector
$$
\Bigl(\frac{w^ndz}{z-a_1}, \ldots, \frac{w^ndz}{z-a_M},\frac{w^ndz}z, \frac{w^ndz}{z-1}\Bigr)
$$
along these very cycles, due to Theorem \ref{thm_conj}, gives us $M+1$ basic elements in the $(M+1)$-dimensional space of triangular solutions (\ref{res_garnier}) of the Schlesinger $(2\times2)$-system in the case $\beta_1=\ldots=\beta_{M+2}=n/2<0$.
These basic solutions are rational, with explicit expressions given by Theorem \ref{thm_ration}, which implies the existence of an $M$-parameter family of algebraic solutions of the corresponding Garnier system.

\begin{theorem}\label{thm_garnier}
For any integer $n<0$, the Garnier system ${\cal G}_M (\theta)$ with parameters
$$
(\theta_1,\ldots,\theta_{M+2},\theta_{\infty})=(\pm n,\ldots,\pm n,-(M+2)n-1)
$$
$($the signs are independent$)$ possesses an $M$-parameter family of algebraic solutions, which can be computed explicitly
by using Theorem \ref{thm_ration}.
\end{theorem}

\begin{example} \label{ex_garnier2}
{\rm Let us illustrate the above theorem by computing two-parameter families of algebraic solutions of bivariate Garnier systems ${\cal G}_2(\pm1,\pm1,\pm1,\pm1, 3)$ (the case of $M=2$, $n=-1$).

Calculating the residues at the three poles $(a_1,0)$, $(a_2,0)$, $(a_3,0)=(0,0)$ of the differentials
$$
\Omega_1=\frac{dz}{w(z-a_1)}, \quad \Omega_2=\frac{dz}{w(z-a_2)}, \quad \Omega_3=\frac{dz}{wz}, \quad \Omega_4=\frac{dz}{w(z-1)},
$$
we have three linear independent vector functions, respectively,
$$
\bigl(b_1^{(j)}(a_1,a_2), b_2^{(j)}(a_1,a_2), b_3^{(j)}(a_1,a_2), b_4^{(j)}(a_1,a_2)\bigr), \qquad j=1,2,3,
$$
where $b_i^{(j)}={\rm res}_{(a_j,0)}\Omega_i$. Using the explicit formula (\ref{rationsol1}) leads to the following expressions:
$$
b_2^{(1)}=\frac1{(a_1-a_2)^2a_1(a_1-1)}, \quad b_3^{(1)}=\frac1{(a_1-a_2)a_1^2(a_1-1)}, \quad b_4^{(1)}=\frac1{(a_1-a_2) a_1(a_1-1)^2},
$$
$$
b_1^{(1)}=-b_2^{(1)}-b_3^{(1)}-b_4^{(1)};
$$
$$
b_1^{(2)}=\frac1{(a_2-a_1)^2a_2(a_2-1)}, \quad b_3^{(2)}=\frac1{(a_2-a_1)a_2^2(a_2-1)}, \quad b_4^{(2)}=\frac1{(a_2-a_1) a_2(a_2-1)^2},
$$
$$
b_2^{(2)}=-b_1^{(2)}-b_3^{(2)}-b_4^{(2)};
$$
$$
b_1^{(3)}=\frac1{a_1^2a_2}, \quad b_2^{(3)}=\frac1{a_2^2a_1}, \quad b_3^{(3)}=-\frac{a_1a_2+a_1+a_2}{a_1^2a_2^2}, \quad b_4^{(3)}=\frac1{a_1a_2}.
$$

Then, like in Example \ref{ex_garnier1}, we consider the polynomial
$$
P_2(z,a_1,a_2)=\bigl(b_4+a_1b_1+a_2b_2\bigr)z^2+\bigl(a_1a_2(b_3+b_4)+a_1(b_2+b_3)+a_2(b_1+b_3)\bigr)z-a_1a_2b_3,
$$
with $b_i=c_1\,b_i^{(1)}+c_2\,b_i^{(2)}+b_i^{(3)}$, which contains two free parameters $c_1,c_2\in{\mathbb C}$ and thus
defines a two-parameter family of algebraic functions, two branches $u_1(a_1,a_2,c_1,c_2)$, $u_2(a_1,a_2,c_1,c_2)$ of which give us
the two-parameter family of algebraic solutions
$$
(u_1,u_2,v_1^{\varepsilon},v_2^{\varepsilon}), \quad v_j^{\varepsilon}(a_1,a_2,c_1,c_2)=
\frac12\Bigl(\frac{-1+\varepsilon_1}{u_j-a_1}+\frac{-1+\varepsilon_2}{u_j-a_2}+\frac{-1+\varepsilon_3}{u_j}+\frac{-1+\varepsilon_4}{u_j-1}\Bigr),
\quad \varepsilon_i\in\{\pm1\},
$$
of the Garnier system ${\cal G}_2\bigl(\varepsilon_1,\varepsilon_2,\varepsilon_3,\varepsilon_4,3\bigr)$.
}
\end{example}

\begin{remark}\label{rmk_6} {\rm Classical solutions of Garnier systems were studied and partially described in \cite{KO}
and in \cite{Maz2}, mainly in terms of the monodromy of a Fuchsian family (\ref{fuchs}) that is governed by
the corresponding Schlesinger $(2\times2)$-system (though, there is no full classification of classical solutions here yet, in contrast
to sixth Painlev\'e equations, whose classical {\it non-algebraic} solutions were classified by H.\,Watanabe \cite{Wat} whereas
the problem of the classification of their {\it algebraic} solutions has been finally closed by O.\,Lisovyy and Yu.\,Tykhyy \cite{LT}).
In particular, if the monodromy of the Fuchsian family is {\it triangular}, the corresponding Garnier system ${\cal G}_M(\theta)$ possesses an $M$-parameter
family of classical solutions expressed {\it via} Lauricella hypergeometric functions (Theorem 6 in \cite{Maz2}). Our case, that of a triangular
Schlesinger system, is certainly included in that context  of triangular monodromy, however, Theorem \ref{thm_5} provides us with an explicit form of
algebraic solutions to particular Garnier systems and, moreover, Theorem \ref{thm_garnier} presents some cases when algebraic solutions of a Garnier
system form an $M$-parameter family.}
\end{remark}

In conclusion of this section let us note that the problem of classification of algebraic solutions to Garnier systems itself is obviously more recent than the analogous one for Painlev\'e VI equations and is still open. Due to G.\,Cousin \cite{Co}, algebraic solutions correspond to finite braid group orbits on the character variety of the $(M+3)$-punctured Riemann sphere ({\it i.~e.}, on the moduli space of its rank two linear monodromy representations). With respect to this correspondence, in the case of a {\it non-degenerate} linear monodromy (that is, neither finite, nor dihedral, nor triangular), algebraic solutions were partially classified in \cite{Di} for an arbitrary $M$ and in \cite{CM} for $M=2$. For a {\it non-abelian triangular} linear monodromy, the classification of Schlesinger isomonodromic $(2\times2)$-families leading to algebraic solutions of Garnier systems, was done in \cite{CoMo}. Note that the monodromy of the triangular Schlesinger isomonodromic families corresponding to algebraic solutions from Theorems \ref{thm_5}, \ref{thm_garnier} is {\it abelian}, similarly to the linear monodromy of the rational solutions to the Painlev\'e VI equations from Theorems \ref{thmrat1}, \ref{thm_rat2}. For a {\it dihedral} linear monodromy, there are families of algebraic solutions obtained in \cite{Gir}, for $M=2$ and in \cite{Kom} for an arbitrary even $M$. Earlier, algebraic solutions of some particular Garnier systems were also proposed in \cite{Ts}, by applying birational canonical transformations to a fixed algebraic solution, without appealing to Schlesinger isomonodromic deformations though.
\medskip

{\bf Acknowledgements.} We thank Vladimir Leksin who had drawn attention of the second author to the paper \cite{DS2}, which has led to
the present work, as well as Irina Goryuchkina for helping us to verify by Maple the solutions of Example \ref{ex_garnier1} (they indeed satisfy the corresponding Garnier systems!).

We thank the anonymous referees for useful suggestions which improved this paper.

V.\,S. is grateful to the Natural Sciences and Engineering Research Council of Canada for the financial support through a Discovery grant.
The research  of V.\,D.  was partially supported by the Serbian Ministry of Education, Science, and Technological
Development, the Science Fund of Serbia and by The University of Texas at Dallas.

\end{document}